\documentclass[a4paper,11pt]{article}
\pdfoutput=1
\usepackage{amssymb}
\usepackage{epsfig}
\usepackage{graphicx}

\makeatletter

\@addtoreset{equation}{section}
\makeatother
\def\be{\begin{equation}}
\def\ee{\end{equation}}
\def\bea{\begin{eqnarray}}
\def\eea{\end{eqnarray}}

\def\({\left(}
\def\){\right)}
\def\<{\left<}
\def\>{\right>}

\def\be{\begin{equation}}
\def\ee{\end{equation}}
\def\bea{\begin{eqnarray*}}
\def\eea{\end{eqnarray*}}
\def\ben{\begin{eqnarray}}
\def\een{\end{eqnarray}}
\def\({\left(}
\def\){\right)}
\def\<{\left<}
\def\>{\right>}
\def\!{\right|}
\def\|{\left|}

\def\[{\left[}
\def\]{\right]}

\def\+{\bar}
\def\mb{\mathbb}

\def\Tr{{\mbox{Tr}}}
\def\D{{\cal{D}}}
\def\L{{\cal{L}}}
\def\t{\widetilde}
\def\A{{\cal{A}}}

\def\J{{\cal{J}}}
\def\N{{\cal{N}}}

\def\F{{\cal{F}}}

\def\H{{\cal{H}}}

\def\L{{\cal{L}}}

\def\F{{\cal{F}}}

\def\Q{{\cal{Q}}}

\begin{document}
\setlength{\unitlength}{1mm}

\begin{titlepage}
\vskip1cm
\begin{flushright}
UOSTP {\tt 150301}
\end{flushright}
\vskip 2.25cm
\centerline{\Large
\bf{The geometric Langlands twist in five and six dimensions}  
}
\vskip 1.25cm \centerline{  Dongsu Bak$^a$ and  Andreas Gustavsson$^{a,b}$}
\vspace{1cm} \begin{center} \it  a) Physics Department,
University of Seoul,  Seoul 130-743, Korea\\
b) School of Physics, Korea Institute for Advanced Study, Seoul 130-012, Korea
\end{center}
\vskip 0.75cm \centerline{(\tt dsbak@uos.ac.kr, agbrev@gmail.com)}
\vspace{2.5cm}
\centerline{\bf Abstract} \vspace{0.75cm} \noindent
Abelian 6d (2,0) theory has SO(5) R symmetry. We twist this theory by identifying the R symmetry group with the SO(5) subgroup of the SO(1,5) Lorentz group. This twisted theory can be put on any five-manifold M, times R, while preserving one scalar supercharge. We subsequently assume the existence of one unit normalized Killing vector field on M, and we find a corresponding SO(4) twist that preserves two supercharges and is a generalization of the geometric Langlands twist of 4d SYM. We generalize the story to non-Abelian gauge group for the corresponding 5d SYM theories on M. We derive a vanishing theorem for BPS contact instantons by identifying the 6d potential energy and its BPS bound, in the 5d theory. To this end we need to perform a Wick rotation that complexifies the gauge field. 
 
\vspace{1.75cm}

\end{titlepage}
\section{Introduction}
It can be interesting to consider Euclidean M5 brane on $\mb{R} \times S^5$ which is conformally equivalent with  $\mb{R}^6$, since this is the conformal boundary of AdS$_7$ and so we can use the AdS/CFT correspondence. Dimensional reduction along $\mb{R}$ is rather tricky since it involves first making $\mb{R}$ compact, and further involves Wick rotating time in the 6d theory. Anyhow, it is expected that this dimensional reduction gives a particular 5d SYM theory on $S^5$. This 5d SYM theory on $S^5$ has been approached using superconformal Killing spinors and Scherk-Schwarz reduction,  for example, in \cite{Hosomichi:2012ek}. It seems that a different, but possibly equivalent, approach to this 5d theory should be also possible, since according to \cite{Bershadsky:1995qy}, when a brane is put on a curved space, the field theory that lives on the brane will be automatically topologically twisted.\footnote{The corresponding question in three dimensions has been recently addressed in \cite{Imbimbo:2014pla}.} Let us assume that we have Lorentzian time, and 
 put M5 brane on $\mb{R} \times M$ where $M$ is a generic five-manifold with no isometries.\footnote{In this paper we will not address the question whether these topologically twisted theories can be realized as brane configutations in M theory. This question has been addressed in \cite{Gauntlett:2000ng}} Then the M5 brane theory should be twisted so that the $SO(5)$ R symmetry is identified with the $SO(5)$ subgroup of the Lorentz group. We then find just one hermitian scalar supercharge $Q$ and we have the supersymmetry algebra
\bea
Q^2 = H
\eea
where $H$ is the Hamiltonian that generates time translation along $\mb{R}$. If the five-manifold has one isometry, then we have an additional bosonic symmetry generator $P$ that generates translation along the Killing direction, and acts on the fields as a Lie derivative. In this case one may expect to find a second hermitian scalar supercharge $Q^{\vee}$, and the supersymmetry algebra
\ben
Q^2 &=& H\cr
(Q^{\vee})^2 &=& H\cr
\{Q,Q^{\vee}\} &=& -2P\label{alg}
\een
There is no reason to expect that this stops here. If $M$ has more isometries, then we may expect to also find more supersymmetries. However we will not consider these situations in this paper. Rather 
we will content ourselves with just one isometry direction on $M$. An interesting special case is when $M = \mb{R} \times M_4$. This  will correspond to the 5d uplift of the geometric Langlands twist of 4d MSYM on $M_4$ \cite{Marcus:1995mq,Kapustin:2006pk,Witten:2011zz}. But the $SO(4)$ twisted theory we obtain can be put on a larger class of five-manifolds. For the existence of a second supersymmetry, we need to assume the existence of one Killing vector field $v^m$ that has unit norm. If we define $v_m = G_{mn} v^n$ and $w_{mn} = \nabla_m v_n - \nabla_n v_m$, we then have the Killing equation $\nabla_m v_n + \nabla_n v_m = 0$ and 
\ben
v^m v_m &=& 1\cr
v^m w_{mn} &=& 0\label{ett}
\een
where the second equation follows directly from the Killing equation and unit normalization condition. On the other hand, the condition for a contact manifold is that we have a globally defined contact form $v_m$ such that  
\ben
\epsilon^{mnpqr} v_m w_{np} w_{qr} \neq 0\label{tva}
\een
at every point. In this case, $v^m = G^{mn} v_n$ that satisfies (\ref{ett}) will be a Reeb vector field of the metric contact manifold, which is uniquely specified by the two conditions (\ref{ett}). But finally, by also requiring that $v^m$ is a Killing vector field, we then  have a (metric) K-contact manifold. For our purposes of obtaining an $SO(4)$ twisted gauge theory, we do not need to assume that $v_m$ defines a contact one-form in the first place. 
The minimal assumptions we have to make are that $v^m$ is a Killing vector field, and that $v^m v_m = 1$. So although our five-manifold can be any K-contact manifold, it appears to us that it can also be something more general than a K-contact manifold.

\subsection{Wick rotation}\label{Wick}
We begin by considering M5 brane on $\mb{R} \times \mb{R}^5$ with Lorentzian time along $\mb{R}$. If we furthermore pick one Killing vector on $\mb{R}^5$, then we can define a Hamiltonian that translates along that Killing direction rather than along Lorentzian time. Let us first compactify $\mb{R}$ to a time-like circle. We assume the metric is 
\bea
ds^2 &=& -T^2 dt^2 + dx^m dx^m
\eea
where $m = 1,2,3,4,5$ are vector indices on $\mb{R}^5$, and we impose the identification
\bea
t &\sim & t + 2\pi
\eea
We perform dimensional reduction along this time-like circle \cite{Hull:2014cxa}. We then obtain 5D SYM Lagrangian 
\ben
S &=& \frac{1}{4\pi^2 T} \int_{\mb{R}^5} d^5 x \(\frac{1}{4} F^{mn} F_{mn} - \frac{1}{2} \partial^m \phi^A \partial_m \phi^A + {\mbox{fermions}}\)\label{L}
\een
on $\mb{R}^5$ that corresponds to Lorentzian M5 brane on $S^1 \times \mb{R}^5$. 

We then like to find the corresponding 5D SYM theory that describes Euclidean M5 brane by Wick rotating the Lorentzian time direction. However, there are no real spinors in Euclidean 6D and no real classical Lagrangian description of Euclidean M5 brane theory. Instead we will define Wick rotation in the 5D theory. For the Lorentzian case, the requirement 
of the Lagrangian is that its bosonic part is real, in order to have a unitary field theory \cite{Witten:1989ip}. Wick rotation should bring the Lorentzian Lagrangian into a positive quantity, by a suitable choice of integration cycle. It should be such that the real part of the action goes to 
plus infinity at the asymptotic infinity of the integration cycle \cite{Witten:2010zr}. 

Since there is no time direction that we can rotate in (\ref{L}) into Euclidean signature (as the five-manifold $\mb{R}^5$ it is already of Euclidean signature) we simply need to look for a suitable integration cycle such that the Lagrangian (\ref{L}) becomes positive definite, and this will then correspond to the Euclidean M5 brane. Since we have the wrong sign kinetic terms for the scalar fields while the gauge field part has the right sign kinetic term, we shall Wick rotate the five scalar fields into the imaginary axis. This Wick rotation also extends to the case of non-Abelian generalization. If we do the Wick rotation on the scalar fields, then all the bosonic terms in the non-Abelian Euclidean Lagrangian become positive definite. 

Let us refine the above discussion slightly. In the Lorentzian path integral we have $\exp i S$ and in the Euclidean path integral we have $\exp - S$. Wick rotation of $t$ can be traded for Wick rotation of $T = -iR$. Then $S\sim \frac{1}{T}\int \L$ will go into $S\sim \frac{i}{R}\int \L$. In other words $\exp iS = \exp \frac{i}{T}\int \L$ will go into $\exp iS = \exp -\frac{1}{R}\int \L$. Then after Wick rotation we need to find a new integration cycle where the action $\int \L$ is positive definite. This is precisely what we did above, where we found the new integration cycle is along the imaginary axis for the scalar fields.

\subsection{Putting the theory on $\mb{R}$ times an arbitrary five-manifold} 
If we like to replace $\mb{R}^5$ with an arbitrary smooth five-manifold $M_5$, then we shall twist the M5 brane theory along $\mb{R}^5$ \cite{Bershadsky:1995qy}. The M5 brane theory has $SO(1,5)$ Lorentz symmetry and $SO(5)$ R symmetry. The twisting amounts to identifying the $SO(5)$ subgroup of $SO(1,5)$, with the R symmetry group $SO(5)$, which in particular means that indices $m$ and $A$ are identified. The scalar fields after the twist thus become a vector field $\phi_m$ on $M_5$. For the spinor field that transforms in the representation $(4,4)$ under $SO(1,5) \times SO(5)$, 
 its representation is organized as
\bea
4 \times 4 &=& 1 \oplus 5 \oplus 10
\eea
after the $SO(5)$ twisting.
We will denote these tensor components as $\psi$, $\psi_m$ and $\psi_{mn} = -\psi_{nm}$ respectively, and expand the spinor in this as
\bea
\psi &=& \frac{1}{2} \(C^{-1} \psi + \gamma^m C^{-1} \psi_m + \gamma^{mn} C^{-1} \psi_{mn}\)
\eea
where $\gamma^m$ and $C$ are respectively  
 gamma matrices and  charge conjugation matrix of $SO(5)$. (The details are given in Appendix A.)  
The scalar supersymmetry parameter $\epsilon_0$ of the twisted theory, corresponds to the bispinor
\bea
\epsilon &=& \frac{1}{2} \epsilon_0 C^{-1}
\eea
We will denote the associated scalar supercharge as $Q$.

If the five-manifold has a Killing vector field $v^m =  \delta^m_5$, then we say that we make an $SO(4)$ twist, rather than an $SO(5)$ twist. We decompose $m = (i,5)$ and we have bosonic fields $(A_{i},
a)$ and $(\phi_{i},\varphi)$ with $a=A_5$ and  $\varphi = \phi_5$. We have fermionic fields as follows. We have two fermionic scalars $\psi,\chi$ where we define $\chi = \psi_5$. We have two fermionic vectors $\psi_{i}$ and $\chi_{i} = \psi_{5i}$, and finally we have one fermionic tensors $\psi_{ij}$ that we may decompose into selfdual and antiselfdual parts. We have a second scalar supersymmetry parameter $\epsilon'_0$ that we define as
\bea
\epsilon' &=& \frac{1}{2} \epsilon'_0 \gamma^5 C^{-1}
\eea
We will denote the associated supercharge as $Q^{\vee}$. This case is the 5d uplift of the geometric Langlands twist in 4d $\N=4$ SYM \cite{Marcus:1995mq,Kapustin:2006pk}. But in 4d we have $SO(6)$ R symmetry that is split as $SO(6) \rightarrow SO(2) \times SO(4)$ and then the $SO(4)$ is identified with the Lorentz group in Euclidean signature. After this geometric Langlands twist, we are left with an $SO(2)$ R symmetry. In 5d  we have no R symmetry left after the $SO(4)$ twist since the original R symmetry is $SO(5)$ instead of $SO(6)$. However we will find that there is a discrete $\mb{Z}_2$ symmetry that acts by exchanging the two supercharges as $Q \rightarrow Q^{\vee}$ and $Q^{\vee} \rightarrow Q$. Let us notice that we do not find that $Q^{\vee} \rightarrow -Q$ with a minus sign. These two supercharges arises from $SO(1,5)$ Lorentz group broken down to $SO(1,1)$ by the $SO(4)$ twist. So our $\mb{Z}_2$ symmetry should be a subgroup of $SO(1,1)$ rather than a subgroup of $SO(2)$.

Let us now consider a six-manifold $M_6$ that is a line bundle over a five-manifold $M$, with the metric 
\ben
ds^2 &=& R^2\(dt+V_m dx^m\)^2 + G_{mn} dx^m dx^n\label{EM}
\een
Here $V_m$ is a graviphoton field on $M$. If we consider a selfdual tensor field in this Euclidean 6d geometry, then this will under dimensional reduction along $t$ down to $M$, give rise to the usual Maxwell term, plus the following graviphoton term  \cite{Witten:2009at, Linander:2011jy}
\bea
S_{graviphoton} &=& \frac{i}{8\pi^2} \int_M V \wedge \Tr \(F \wedge F\)
\eea
The factor of $i$ that appears in the graviphoton term comes from the $i$ of the selfduality relation $*H = iH$ in Euclidean signature. The factor of $i$ is also necessary to have $\exp \frac{i}{8\pi^2} \int V\wedge \Tr(F\wedge F)$ in the path integral, as it should be. One way to argue for that is by dimensionally reducing to four dimensions where this becomes the theta term that is $2\pi$ periodic.

\subsection{Reflection positivity}
There is another way to argue for the factor of $i$ in the graviphoton term in Euclidean signature. This is criterion that the action shall be reflection positive in order to have an analytic continuation to Lorentzian theory. Since we treat Euclidean time and Euclidean space directions differently upon dimensional reduction along Euclidean time, we will use a restricted definition of reflection positivity in the dimensionally reduced theory, saying that the action of the dimensionally reduced Euclidean theory shall be invariant under time reversal combined with complex conjugation. If time and space directions are on equal footing in an Euclidean theory, then we can replace time reversal with parity on the Euclidean space.

The Euclidean 6d metric (\ref{EM}) is invariant under the Euclidean time reversal $t\rightarrow -t$ provided  that the graviphoton transforms as 
\bea
V_m &\rightarrow & -V_m
\eea
Moreover, since the gauge field $A_m$ originates from the two-form potential $B_{mt}$ in the 6d tensor multiplet theory, whereas $\phi_m$ origines from scalar fields $\phi^A$, we should have 
\bea
A_m &\rightarrow & - A_m\cr
\phi_m &\rightarrow & \phi_m
\eea
under time reversal $t\rightarrow -t$. We then see that $i V \wedge \Tr(F \wedge F) \rightarrow -iV\wedge \Tr(F \wedge F)$ under time reversal. But then complex conjugation brings it back to $iV \wedge \Tr(F\wedge F)$ so it will be reflection positive. Let us notice that by integration by parts, we can write the graviphoton term in the form $i dV \wedge Q(A)$ where $Q(A)$ is the Chern-Simons functional, defined such that $dQ(A) = F \wedge F$. In \cite{Cordova:2013cea,Lee:2013ida,Yagi:2013fda} a complex CS term was found in $SO(3)$ twisted 5d SYM. It is natural to think that this complex CS term arises from a complexified version of a  graviphoton term in 5d. If that is the case, then by the above transformation rules of the graviphoton and the complex gauge field $\A_m = A_m + i \phi_m \rightarrow - \bar\A_m$ under time reversal, this Chern-Simons term will be reflection positive when the complex CS level $q = k + i \sigma$ is real, which means that $\sigma$ has to be purely imaginary (and $k$ has to be an integer). We notice that the possibility of having a second branch where $\sigma$ is purely imaginary instead of real was found already in \cite{Witten:1989ip}.

\section{Abelian 6d $(2,0)$ theory}
There is no twist of 6d $(2,0)$ theory that can make it a fully topological field theory, since the R-symmetry group $SO(5)$ is smaller than the Lorentz group $SO(1,5)$. The best we can do, is to do a partial topological twist by identifying the $SO(5)_L$ subgroup of $SO(1,5)$ with the $SO(5)_R$ R-symmetry. To describe this twisted theory, we start from untwisted theory on flat space. We then perform the twist, still on flat space, and finally we find that the resulting theory can be put on a Lorentzian six-manifold of the form $\mb{R} \times M$, where $M$ can be any five-manifold, while preserving  one supercharge $Q$. This supercharge will square to the Hamiltonian which generates time translation along $\mb{R}$. 

The abelian 6d $(2,0)$ tensor multiplet theory on flat ${\mb{R}}^{1,5}$ with Lorentzian metric $\eta_{MN}=$ diag$(-1,1,1,1,1,1)$, can be captured by the action
\ben
S &=& \int d^6 x\(-\frac{1}{24} H^{MNP} H_{MNP} - \frac{1}{2} \partial^M \phi^A \partial_M \phi_A + \frac{i}{2} \bar\psi \Gamma^M \partial_M \psi\)\label{action}
\een
Here $H_{MNP}$ is a non-selfdual tensor field, $\phi^A$ are five scalar fields, and $\psi$ are fermions  that are subject to a 11d Majorana condition and 6d Weyl condition. In Lorentzian signature, the tensor field can be separated into real selfdual and real antiselfdual parts,
\bea
H_{MNP} &=& H_{MNP}^+ + H_{MNP}^-
\eea
and it is only the selfdual piece that belongs to the tensor multiplet. Nevertheless, the action (\ref{action}) is invariant under the 6d $(2,0)$ on-shell supersymmetry variations
\ben
\delta B_{MN} &=& i\bar\epsilon\Gamma_{MN}\psi\cr
\delta \phi^A &=& i\bar\epsilon\Gamma^A\psi\cr
\delta \psi &=& \frac{1}{12}\Gamma^{MNP}\epsilon H^+_{MNP} + \Gamma^M\Gamma_A\epsilon \partial_M\phi^A\label{6dvariations}
\een
On-shell means that these variations form a closed algebra on-shell, but the action is invariant under these variations without using any equations of motion. If we would use equations of motion then the action will be stationary and hence invariant under any variations of course. The supersymmetry variation of the field strength can be expressed as 
\bea
\delta H_{MNP} &=& \frac{i}{2} \bar\epsilon \Gamma^Q \Gamma_{MNP} \partial_Q \psi
\eea
by using the fermionic equation of motion $\Gamma^Q \partial_Q \psi = 0$. This together with the 6d Weyl condition of the spinor $\psi$, shows
 that the antiselfdual part is invariant, $\delta H_{MNP}^- = 0$, under the supersymmetry variation. Nevertheless, we need to keep the $H^-_{MNP}$ components in the action as spectator field components, since otherwise we have troubles to write a covariant action in a simple way. The supercurrent is given by 
\bea
j^M &=& -\frac{1}{12}\bar\epsilon \Gamma^{RST} \Gamma^M \psi H_{RST}^+ 
- \bar\epsilon \Gamma^A \Gamma^N \Gamma^M \psi \partial_N \phi^A
\eea

\subsection{The $SO(5)$ twist}
We will now perform the $SO(5)$ twist that amounts to identifying the $SO(5)_R$ R symmetry group with the $SO(5)_L$ subgroup of the $SO(1,5)$ Lorentz group. We denote spinor components as $\psi^{\alpha I i}$ where $\alpha$ is 4-component spinor index of $SO(5)\subset SO(1,5)$, $I = +,-$ is the spinor index that corresponds to the 6d chirality of the spinor, and $i$ is the 4-component spinor index of $SO(5)$ R symmetry. We denote vector indices of $SO(1,5)$ as $M = (0,m)$ where $m$ is vector index of $SO(5)_L$. Our gamma matrix conventions are collected in Appendix \ref{spinor}. After the twist, we have reduced the global symmetry of the theory down to the diagonal subgroup $SO(5)'=$diag$[\, SO(5)_R \times SO(5)_L]$, which will be the twisted Lorentz symmetry. We write the scalar fields $\phi^A$ as $\phi_m$, and we identify the spinor indices $\alpha$ and $i$, both transforming as 4-component spinor under the twisted Lorentz symmetry. In the Appendix \ref{Majorana}, we show how the 6d Weyl spinor that is subject to the 11d Majorana condition, is expanded into sixteen real components $\psi$, $\psi_m$ and $\psi_{mn}=-\psi_{nm}$.

After the twist, the Lagrangian becomes  
\bea
\L &=& \L_B+\L_{\phi} + \L_1+\L_2
\eea
where
\bea
\L_B &=& \frac{1}{8} H_0{}^{mn} H_{0mn} - \frac{1}{24} H^{mnp} H_{mnp} \cr
\L_{\phi} &=& - \frac{1}{2} \partial^m \phi^n \partial_m \phi_n\cr
\L_1 &=&  i \psi^m \partial_m \psi + 2i \psi^{mn} \partial_m \psi_n - \frac{i}{2} \epsilon^{mnpqr} \psi_{mn} \partial_p \psi_{qr}\cr
\L_2 &=& \frac{1}{2} \partial_0 \phi^m \partial_0 \phi_m - \frac{i}{2}\(\psi \partial_0 \psi + \psi^m \partial_0 \psi_m + 2 \psi^{mn} \partial_0 \psi_{mn}\)
\eea
We can also introduce one auxiliary scalar field $\phi$ and make the following substitution
\ben
\L_{\phi} &=& - \frac{1}{4} \phi^{mn}\phi_{mn} + \frac{1}{2} \phi^2 - \phi \partial^m \phi_m\label{replace}
\een
where we define
\bea
\phi_{mn} &=& \partial_m \phi_n - \partial_n \phi_m
\eea
Integrating out the auxiliary scalar, we find its value as $\phi = \partial^m \phi_m$, and by plugging this back into the Lagrangian we find 
\bea
\L_{\phi} &=& -\frac{1}{4} \phi^{mn} \phi_{mn} - \frac{1}{2} (\partial^m \phi_m)^2
\eea
which is a rewriting of the original term. We can put this Lagrangian on a curved six-manifold 
 of the form $\mb{R}\times M$ where time is along $\mb{R}$, provided that we understand that indices are now being raised by the inverse metric $G^{mn}$ of the five-manifold $M$. The ordinary derivatives should be replaced by covariant derivatives $\nabla_m$, but in antisymmetric combinations this is not necessary since $\nabla_m \psi_n - \nabla_n \psi_m = \partial_m \psi_n - \partial_n \psi_m$. 

Let us now return to $\L_{\phi}$ once more. If we introduce a one-form as $\phi = \phi_m dx^m$, and an inner product of two p-forms $a$ and $b$ as $(a,b) = \int_M d^5 x \sqrt{G} a\wedge *b$, then this term can be expressed as
\bea
\int_{\mb{R}}  dt\int_M d^5 x \sqrt{G} \L_{\phi} &=& \int_{\mb{R}} dt \(-\frac{1}{2}(d\phi,d\phi) - \frac{1}{2} (d^{\dag}\phi,d^{\dag}\phi)\) = -\int_{\mb{R}} dt \frac{1}{2}(\phi,\triangle \phi)
\eea
where $\triangle = d d^{\dag} + d^{\dag} d$ is the Laplace operator acting on one-forms on $M$. This form of the Lagrangian should be useful for quantization since it amounts to computing the functional determinant of $\triangle$ on $M$. 

It is straightforward to obtain the supersymmetry variations that correspond to the supersymmetry parameter
\bea
\epsilon^{\alpha-i} &=& \frac{1}{2} \epsilon C^{\alpha i}
\eea
in the untwisted theory. Here $\epsilon$ is a Grassmann odd scalar supersymmetry parameter. The resulting twisted supersymmetry variations are
\ben
\delta_{\epsilon} B_{mn} &=& 2i\epsilon\psi_{mn}\cr
\delta_{\epsilon} B_{m0} &=& -i\epsilon\psi_m\cr
\delta_{\epsilon} \phi_m &=& -i\epsilon \psi_m\cr
\delta_{\epsilon} \phi &=& -i\epsilon\partial_0 \psi\cr
\delta_{\epsilon} \psi &=& -\epsilon \phi\cr
\delta_{\epsilon} \psi_m &=& -\epsilon \partial_0 \phi_m\cr
\delta_{\epsilon} \psi_{mn} &=& \frac{1}{2}\epsilon H_{0mn}^+ - \frac{1}{2} \epsilon \phi_{mn}\label{6dQvariations}
\een
Here we define
\ben
H_{0mn}^{\pm} &=& \frac{1}{2} \(H_{0mn} \mp \frac{1}{6} \epsilon_{mnpqr} H^{pqr}\)\pm C_{mn}\label{SD}
\een
and from the untwisted theory we have $C_{mn} = 0$. But we include this extra term here because it will be required when we put the theory on a five-manifold that has one isometry direction, in order to have enhancement to two supercharges.  

In a similar way we can obtain the supercurrent in the twisted theory from the supercurrent of the untwisted theory. The result is 
\bea
j^0 &=& \psi^{mn} (H_{0mn}^+ - \phi_{mn}) - \psi \phi - \psi^m \partial_0 \phi_m\cr
j^m &=& -(H_{0}^{+mn} - \phi^{mn})\psi_n + \psi^m \phi + \frac{1}{2} \epsilon^{mnpqr} (H_{0np}^+ -\phi_{np}) \psi_{qr} - 2 \psi^{mn} 
\partial_0 \phi_n
\eea
If we define $\delta$ as the anticommuting supersymmetry variation we get from the commuting supersymmetry variation $\delta_{\epsilon}$ by removing the fermionic parameter $\epsilon$, then $j^0$ can be neatly expressed as
\ben
j^0 &=& 2\psi^{mn} \delta \psi_{mn}+ \psi^m \delta \psi_m + \psi \delta \psi \label{j0}
\een

\subsection{The $SO(4)$ twist}
Let us return to flat $\mb{R}^{1,5}$ and let us now instead view this space as $\mb{R}^{1,1} \times \mb{R}^4$. We may then twist the theory by identifying the $SO(4)\subset SO(1,5)$ that rotates in the $\mb{R}^4$ part of the space, with an $SO(4)$ subgroup of the $SO(5)$ R symmetry. We write spacetime indices as $M =(0,i,5)$ where $i=1,2,3,4$ is vector index in $\mb{R}^4$. We now find a second scalar supersymmetry parameter as
\bea
\epsilon^{\alpha - i} &=& \frac{1}{2} \epsilon (\gamma_5)^{\alpha i}
\eea
where we identify the two $SO(4)$ Dirac spinor indices $\alpha$ and $i$ (with apologies for using the same index $i$ for both spinor and vector on $\mb{R}^4$, but after the twist is done, no spinor indices will appear anywhere). We then obtain a second set of twisted supersymmetry variations as
\bea
\delta^{\vee} \phi_5 &=& -i\epsilon \psi\cr
\delta^{\vee} \phi_i &=& -2i\epsilon \psi_{5i}\cr
\delta^{\vee} B_{i0} &=& +2i\epsilon \psi_{5i}\cr
\delta^{\vee} B_{50} &=& -i\epsilon \psi\cr
\delta^{\vee} B_{5i} &=& -i\epsilon \psi_i\cr
\delta^{\vee} B_{ij} &=& -i\epsilon \epsilon_{ijkl} \psi^{kl}\cr
\delta^{\vee} \psi &=& -\epsilon \partial_0 \phi_5\cr
\delta^{\vee} \psi_i &=& \epsilon (H_{0i5}^+ - \partial_5 \phi_i -\partial_i \phi_5)\cr
\delta^{\vee} \psi_5 &=& \epsilon (\partial_i \phi^i - \partial_5 \phi^5)\cr
\delta^{\vee} \psi_{ij} &=& -\frac{1}{4}\epsilon \epsilon_{ijkl} (H_{0}^{+kl} + \phi^{kl})\cr
\delta^{\vee} \psi_{5i} &=& -\frac{1}{2}\epsilon \partial_0 \phi_i
\eea
Again we can take these variations off-shell by introducing an auxiliary scalar field $\phi^{\vee}$ and let
\bea
\delta^{\vee} \chi &=& -\epsilon \phi^{\vee}\cr
\delta^{\vee} \phi^{\vee} &=& - i\epsilon \partial_0 \chi
\eea
These are related to the original twisted supersymmetries by a $\mb{Z}_2$ transformation
\bea
B_{MN} &\leftrightarrow &B_{MN}\cr
\psi_{ij} &\leftrightarrow & -\frac{1}{2}\epsilon_{ijkl} \psi^{kl}\cr
\chi_i &\leftrightarrow & -\frac{1}{2}\psi_i\cr
\chi &\leftrightarrow & \psi\cr
\phi_i &\leftrightarrow & -\phi_i\cr
\phi &\leftrightarrow & \phi^{\vee}\cr
\eea
In particular then, we have the on-shell values
\bea
\phi &=& \partial_m \phi^m\cr
\phi^{\vee} &=& -\partial_i \phi^i +\partial_5 \phi^5
\eea
As we mentioned already in Introduction, we can generalize this $SO(4)$ twist to five-manifolds that 
have at least one unit normalized Killing vector $v^m$. We define the corresponding one-form and curvature two-form as
\ben
v_m &=& G_{mn} v^n\label{defs1}\\
w_{mn} &=& \partial_m v_n - \partial_n v_m\label{defs2}
\een
and we thus assume the following conditions 
\ben
\nabla_m v_n + \nabla_n v_m &=& 0\label{basicassumption1}\\
G_{mn} v^m v^n &=& 1\label{basicassumption2}
\een
By using the assumptions (\ref{basicassumption1}) and (\ref{basicassumption2}) together with the definitions (\ref{defs1}) and (\ref{defs2}), we obtain the following relations,
\ben
w_{mn} v^n &=& 0\label{contact}\\
-\frac{1}{2}\nabla^m w_{mn} &=& R_{mn}v^m
\label{kcurv}\\
R_{mn} v^m v^n &=& \frac{1}{4} w_{mn} w^{mn}\label{curv}
\een
We use the Killing vector to define the trace parts
\bea
\varphi &=& v^m \phi_m\cr
\chi &=& v^m \psi_m\cr
\chi_m &=& v^n \psi_{nm}
\eea
and then we separate the fields into traceless and trace parts,
\bea
\phi_m &=& \phi'_m + v_m \varphi\cr
\psi_m &=& \psi'_m + v_m \chi\cr
\psi_{mn} &=& \psi'_{mn} + v_m \chi_n - v_n \chi_m
\eea
We use prime to indicate a traceless field\footnote{$\chi_m$ is also traceless but we denote it without prime for the notational simplicity.}. 
We now declare that the $\mb{Z}_2$ transformations shall act as
\bea
B_{MN} &\leftrightarrow & B_{MN}\cr
\psi'_{mn} &\leftrightarrow & -\frac{1}{2} \epsilon_{mnpqr} \psi'^{pq} v^r\cr
\chi_m &\leftrightarrow & -\frac{1}{2} \psi'_m\cr
\chi &\leftrightarrow & \psi\cr
\phi'_m &\leftrightarrow & -\phi'_m\cr
\varphi &\leftrightarrow &\varphi
\eea
which implies that the $\mb{Z}_2$ symmetry acts on the original fields as
\bea
B_{MN} &\rightarrow & B_{MN}\cr
\psi_{mn} &\rightarrow & -\frac{1}{2}\epsilon_{mnpqr} \psi^{pq} v^r - \frac{1}{2} (v_m \psi_n - v_n \psi_m)\cr
\chi_m &\rightarrow & -\frac{1}{2} (\psi_m - v_m \chi)\cr
\psi_m &\rightarrow & -2\chi_m + v_m \psi\cr
\chi &\rightarrow & \psi\cr
\phi_m &\rightarrow & -\phi_m + 2 v_m \varphi\cr
\varphi &\rightarrow & \varphi\cr
\phi &\rightarrow & \phi^{\vee}
\eea
Acting with these transformations on the first supersymmetry variations, we obtain a second set of supersymmetry variations as
\bea
\delta^{\vee} \phi_m &=& -2i \chi_m - i v_m \psi\cr
\delta^{\vee} B_{m0} &=& 2i \chi_m - i v_m \psi\cr
\delta^{\vee} B_{mn} &=& -i \epsilon_{mnpqr} \psi^{pq} v^r - i v_m \psi_n + i v_n \psi_m\cr
\delta^{\vee} \psi &=& -\partial_0 \varphi\cr
\delta^{\vee} \psi_m &=& -v^n H^+_{0nm} - v^n \nabla_n \phi_m - \partial_m \varphi + v_m \nabla_n \phi^n - \frac{1}{2} w_{mn} \phi^n\cr
\delta^{\vee} \psi_{mn} &=& -\frac{1}{4} \epsilon_{mnpqr} (H_0^{+pq} + \phi^{pq} - 2w^{pq} \varphi)v^r - \frac{1}{2} v_m \partial_0 \phi_n + \frac{1}{2} v_n \partial_0 \phi_m
\eea
or by keeping the auxiliary scalar, we have
\bea
\delta^{\vee} \phi^{\vee} &=& -i \partial_0 \chi\cr
\delta^{\vee} \chi &=& -\phi^{\vee}\cr
\delta^{\vee} \psi_m &=& -v_m \phi^{\vee} - v^n H^+_{0nm} - \(v^n \nabla_n \phi_m + w_{mn} \phi^n\) - v^n \nabla_m \phi_n + 2 v_m v^n \nabla_n \varphi
\eea
We can also obtain the variations of the trace parts from the above,
\bea
\delta^{\vee} \varphi &=& -i\psi\cr
\delta^{\vee} \chi &=& - 2 v^m \nabla_m \varphi + \nabla_m \phi^m\cr
\delta^{\vee} \chi_m &=& \frac{1}{2} v_m\partial_0 \varphi - \frac{1}{2} \partial_0 \phi_m
\eea

\subsection{The supersymmetric action}
From the closure computation in Appendix \ref{C}, we can partly deduce the form on the fermionic equations of motion. Requiring these equations are closed under $\mb{Z}_2$ transformation, we find the following fermionic equations of motion
\ben
\nabla_m \psi^m &=& \partial_0 \psi\cr
\nabla_a \psi_b - \nabla_b \psi_a - \epsilon_{abmnp} \nabla^m \psi^{np} - w_{ab}\chi &=& 2\partial_0 \psi_{ab}
\cr
\nabla_m \psi + 2 \nabla^n \psi_{nm} + v_m w_{np} \psi^{np} +\frac{1}{2}\psi^p \Omega_{pm} &=& \partial_0 \psi_m\label{fab}
\een
where we define
\bea
\Omega_{mn} &=& \frac{1}{2} \epsilon_{mnabc} w^{ab}v^c
\eea
Varying with respect to the first supercharge we get respectively, the following bosonic equations of motion
\bea
\partial_0 \(\phi - \nabla_m \phi^m\) &=& 0\cr
\partial_0 ( H_0{}^{mn}+\frac{1}{6} \epsilon^{mnabc}H_{abc}-w^{mn}\varphi )&=& 0\cr
\(-\partial^0 \partial_0 - \nabla^n \nabla_n\) \phi_m + R_{mn} \phi^n + \frac{1}{2} v_m w_{np} H_0{}^{np}&&\cr
- v_m w^{ab} \nabla_a \phi_b + \nabla^p(w_{pm} \varphi)&&\cr
+ \nabla^p (H_{0pm}-w_{pm}\varphi) &&\cr
-\frac{1}{2} \partial_0 \phi^p \epsilon_{pmabc}w^{ab} v^c
&=& 0
\eea
The fermionic equations of motion (\ref{fab}) can be integrated to a Lagrangian
\bea
\L_F &=& -i \nabla_m \psi^m \psi - 2i \nabla_m \psi_n \psi^{mn}\cr
&& - \frac{i}{2} \epsilon^{mnabc} \psi_{mn} \nabla_a \psi_{bc}\cr
&& - i w^{mn} v^p \psi_{mn} \psi_p - \frac{i}{2} \Omega_{mn} \psi^m \psi^n\cr
&& - \frac{i}{2} \psi \partial_0 \psi - \frac{i}{2} \psi^m \partial_0 \psi_m - i \psi^{mn} \partial_0 \psi_{mn}
\eea
By adding the following bosonic Lagrangian
\bea
\L_B &=& \frac{1}{8} H_0{}^{mn} H_{0mn} - \frac{1}{24} H^{mnp} H_{mnp} \cr
&& -\frac{1}{4} \phi^{mn} \phi_{mn}  - \frac{1}{2} (\nabla^m \phi_m)^2 \cr
&& - \frac{1}{2} w^{mn} \(H_{0mn}^+ - \phi_{mn}\) \varphi + \frac{1}{8} w^{mn} w_{mn} \varphi^2 \cr
&&+ \frac{1}{2} \partial_0 \phi^m \partial_0 \phi_m - \frac{1}{2} \Omega_{mn} \phi^m \partial_0 \phi^n
\eea
we find that the combined Lagrangian $\L = \L_F + \L_B$ becomes invariant under $Q$ supersymmetry variations, up to boundary terms. Furthermore, the Lagrangian has $\mb{Z}_2$ symmetry and is therefore invariant under $Q^{\vee}$ supersymmetry 
at the same time.

We notice that the gauge field part of this Lagrangian, combined with the mass term for $\varphi$,
\bea
\L_{B_{MN}} &=& \frac{1}{8} H_0{}^{mn} H_{0mn} - \frac{1}{24} H^{mnp} H_{mnp}\cr
 && - \frac{1}{2} w^{mn} \(\frac{1}{2} H_{0mn} - \frac{1}{12} \epsilon_{mnabc} H^{abc} + \frac{1}{4} w_{mn} \varphi\)\varphi
\eea
can be written in the form
\bea
\L_{B_{MN}} &=&  \frac{1}{24} \(3 (dB)_{0mn}^2 - \(dB + C\)_{mnp}^2\) + \frac{1}{6} \epsilon^{abcmn} C_{abc} (dB)_{0mn} 
\eea
where we define
\bea
C_{mnp} &=& -\frac{1}{2} \epsilon_{mnpab} w^{ab} \varphi\cr
C_{0mn} &=& 0
\eea
This can be compactly rewritten as
\ben
\L_{B_{MN}} &=& -\frac{1}{4} \(|dB + C|^2 - 2 dB \wedge C\)\label{BMN}
\een
Thus we can identify $C$ as the background three-form potential of 11d supergravity that couples to the M5 brane selfdual tensor gauge field. 

\subsection{The Hamiltonian from the Noether procedure}
The conserved current associated with time translation that we obtain by applying the Noether procedure, is given by
\bea
(\Delta t) H &=& -\sum_{\Phi}\delta \Phi \frac{\partial \L}{\partial \partial_0 \Phi} - (\Delta t) \L
\eea
where the sum is over all fields in the theory, and $\Delta t$ is a small constant parameter. Here $\delta \Phi$ is the variation of 
the field under corresponding time translation. To get a gauge covariant expression, we add an appropriate gauge variation. We thus take 
\bea
\delta B_{mn} &=& -(\Delta t) H_{0mn}\cr
\delta \phi_m &=& -(\Delta t) \, \partial_0 \phi_m
\eea
and we compute conjugate momenta to $B_{mn}$ and $\phi_m$ as
\bea
E^{mn} &=& \frac{1}{4} H_0{}^{mn} - \frac{1}{4} w^{mn} \varphi\cr
p^m &=& \partial_0 \phi^m + \frac{1}{2} \Omega^{mn} \phi_n
\eea
The conjugate momentum to $B_{m0}$ is $E^{m0} = 0$ and the consistency condition of this constraint, $\partial_0 E^{m0} = 0$, gives us the 6d 
Gauss law 
\ben
\nabla^m \(H_{0mn} - w_{mn} \varphi\) &=& 0\label{Gaussin6d}
\een
which is the $B_{m0}$ equation of motion. Using the selfduality equation $H^-_{0mn} = 0$, we obtain the (bosonic part of the) Hamiltonian as
\ben
H &=& \frac{1}{4} H_{0mn}H_0{}^{mn} + \frac{1}{4} \phi^{mn} \phi_{mn} + \frac{1}{2} (\nabla^m \phi_m)^2 - \frac{1}{2} w^{mn} \phi_{mn} \varphi + \frac{1}{2} (\partial_0 \phi_m)^2\label{HNoether}
\een
This Hamiltonian seems to be indefinite for $w_{mn}$ nonzero, but we shall show below that it is indeed positive semidefinite.

\subsection{The Hamiltonian from the supercurrent} 
We have two supercharges $Q$ and $Q^{\vee}$ and we can form any linear combination of these
\bea
\Q &=& a Q + b Q^{\vee}
\eea
with real parameters $a$ and $b$ and it will again be a supersymmetry of the theory. We can compute its square. Using the supersymmetry algebra (\ref{alg}), we get
\bea
\Q^2 &=& \(a^2 + b^2\) H - 2ab P
\eea
Since $\Q$ is hermitian when $a$ and $b$ are real, we also have that the left-hand side is greater than or equal to zero. Hence we have a BPS bound
\bea
H &\geq & \frac{2|ab|}{a^2+b^2}|P|
\eea
We have the supercurrent component
\bea
j^0 &=& \psi \delta \psi + \psi^m \delta \psi_m + 2\psi^{mn} \delta \psi_{mn}
\eea
The Hamiltonian is given by 
\bea
2H &=& \delta j^0
\eea
where $\delta$ is defined as the the supersymmetry variation with the supersymmetry parameter removed so that $\delta$ is anticommuting. This splits into a bosonic part $H_B$ and a fermionic part $H_F$ as $H = H_B + H_F$ where
\bea
2 H_B &=& \delta \psi \delta \psi + \delta \psi^m \delta \psi_m + 2 \delta \psi^{mn} \delta \psi_{mn}\cr
2 H_F &=& -\psi \delta^2 \psi - \psi^m \delta^2 \psi_m - 2 \psi^{mn} \delta^2 \psi_{mn}
\eea
Here the minus sign in $H_F$ comes from that $\delta$ is anticommuting. Explicitly we get
\bea
2 H_B &=& \partial_0 \phi^m \partial_0 \phi_m +  \frac{1}{2} \( H_0{}^{+mn} -\phi^{mn}\) \(H_{0mn}^+ -\phi_{mn}\) + (\nabla^m\phi_m)^2 \cr
2 H_F &=& -2i \psi^m \nabla_m \psi - 4i \psi^{mn} \nabla_m \psi_n + i \epsilon_{mnabc} \psi^{mn} \nabla^a \psi^{bc} \cr
&& - \frac{i}{2} \epsilon_{pmabc} w^{ab} v^c \psi^m \psi^p - 2i w^{mn} \chi \psi_{mn}
\eea
To obtain $H_F$ we have used the fermionic equations of motion (\ref{fab}) along with the closure relation $\delta^2 = i\partial_0$.

By making a $\mb{Z}_2$ transformation of $j^0$, we get
\bea
j^{0\vee} &=& \psi \delta^{\vee} \psi + \psi^m \delta^{\vee} \psi_m + 2 \psi^{mn} \delta^{\vee} \psi_{mn}
\eea
We can now also form the linear combination 
\bea
\J^0 &=& aj^0 + b j^{0\vee}
\eea
very easily. We then can obtain $\Q^2$ by computing
 $(a \delta + b \delta^{\vee})\J^0$, from which we can extract $H$ and $P$. If we put the fermions to zero and if we define
\bea
V_{mn}(t) &=& \(a\delta + b \delta^{\vee}\) \psi_{mn}\cr
V_m(t) &=& \(a\delta + b \delta^{\vee}\) \psi_m\cr
V(t) &=& \(a\delta + b \delta^{\vee}\) \psi
\eea
then we have
\bea
(a \delta + b \delta^{\vee})\J^0 &=& 2V^{mn}(t) V_{mn}(t) + V^m(t) V_m(t) + V(t)^2
\eea
and we get the identity
\bea
 \frac{1}{2(a^2+b^2)}\int d^5 x \sqrt{G} \( 
 2V^{mn}(t) V_{mn}(t) + 
V^m(t) V_m(t) + 
V(t)^2\) &=&  
H - \frac{2ab}{(a^2 + b^2)} P
\eea
By defining 
\bea
a &=& u+v\cr
b &=& u-v
\eea
and 
\bea
t &=& \frac{v}{u}
\eea
this relation seems to be a generalization of equation (3.33) in \cite{Kapustin:2006pk}. In particular we can find the combinations
\bea
\frac{ab}{a^2+b^2} &=& \frac{1}{2} \frac{t^{-1} - t}{t^{-1} + t}\cr
\frac{u^2}{a^2 + b^2} &=& \frac{1}{2} \frac{t^{-1}}{t^{-1} + t}\cr
\frac{v^2}{a^2 + b^2} &=& \frac{1}{2} \frac{t}{t^{-1} + t}
\eea
which agree with the coefficients that were chosen in \cite{Kapustin:2006pk}. 

The BPS bound can now be saturated by putting fermions to zero, and by solving the supersymmetric equations $V_{mn}(t) = 0$, $V_m(t) = 0$ and $V(t) = 0$.  However, solving just these equations alone will not be enough since we also have to satisfy the 6d Gauss law constraint (\ref{Gaussin6d}). We extract from $\delta j^0$ that the bosonic part of the Hamiltonian is given by 
\bea
H &=& \frac{1}{2} (\partial_0 \phi_m)^2 + \frac{1}{4} \(H_{0mn}^+ - \phi_{mn}\)^2 + \frac{1}{2} \(\nabla^m \phi_m\)^2
\eea
When we expand out the square in the second term we find a term $-\frac{1}{2} H^+_{0mn} \phi^{mn}$. By integrating by parts, we can write this term as $(\nabla^m H_{0mn}^+) \phi^n$. We next use the Gauss law plus selfduality in the form $\nabla^m H_{0mn}^+ = \nabla^m (w_{mn} \varphi)$, and perform a subsequent integration by parts. We end up with $- \frac{1}{2} w^{mn} \varphi \phi_{mn}$, agreeing with (\ref{HNoether}). To proceed further, we separate $\phi_m = \phi'_m + v_m \varphi$. We then find that we can write the Hamiltonian in the following form
\ben
H &=& \frac{1}{2} (\partial_0 \phi'_m)^2 + \frac{1}{2} (\partial_0 \varphi)^2 + \frac{1}{4} \(H_{0mn}^+  - w_{mn} \varphi - v_n \nabla_m \varphi + v_m \nabla_n\varphi\)^2\cr
&& + \frac{1}{4} \(\nabla_m \phi'_n - \nabla_n \phi'_m\)^2 + \frac{1}{2} \(\nabla^m \phi'_m\)^2 + \frac{1}{2} \(v^m \nabla_m \varphi\)^2\label{6dH}
\een
In this form it is manifest that the Hamiltonian is invariant under the $\mb{Z}_2$ symmetry $\phi'_m \rightarrow -\phi'_m$ and $\varphi \rightarrow \varphi$.

Taking $ab=0$, we find the BPS equations 
\bea
H_{0mn}^+  - w_{mn} \varphi - v_n \nabla_m \varphi + v_m \nabla_n\varphi &=& 0\cr
\nabla_m \phi'_n - \nabla_n \phi'_m &=& 0\cr
\nabla^m \phi'_m &=& 0\cr
v^m \nabla_m \varphi &=& 0\cr
\partial_0 \phi_m &=& 0
\eea
while setting all the fermionic fields  to zero. On these solutions the Hamiltonian is zero. Applying Gauss law on the first equation, we get
\bea
\nabla^m \nabla_m \varphi &=& 0
\eea

\section{Topological 5d SYM on an arbitrary five-manifold}
So far we have considered two different twists of the 6d theory, on $\mb{R} \times M$ where $M$ for the first $SO(5)$ twist was arbitrary and for the second $SO(4)$ twist it had a unit normalized Killing vector $v^m$. Working in 6d has the advantage that the Hamiltonian has a clear interpretation. If we perform `dimensional reduction' by putting time derivatives to zero, then we get an `Euclidean' 5d SYM theory on $M$. Since we no longer have a physical time direction, the Hamiltonian from this 5d point of view becomes obscure. Of course the disadvantage is that in 6d we do not know the nonabelian generalization. 

We will now obtain the Euclidean 5d theories that correspond to Lorentzian 6d theories for $SO(5)$ and $SO(4)$ twists respectively. We thus reduce along Lorentzian time. We also generalize to nonabelian gauge group under which all the fields transform in the adjoint representation. In 5d we have a vector potential $A_m$. By the $SO(5)$ twist, we also have a scalar vector field $\phi_m$. We find it convenient to define the combinations  
\bea
\A_m &=& A_m - \phi_m\cr
\bar\A_m &=& A_m + \phi_m
\eea
By looking at the 6d supersymmetry variations (\ref{6dQvariations}), it is easy to guess that that we shall have the following 5d, nonabelian, supersymmetry variations
\bea
\delta \bar\A_m &=& -2i\epsilon \psi_m\cr
\delta \A_m &=& 0\cr\
\delta \phi &=& 0\cr
\delta \psi_{mn} &=& \frac{1}{2}\epsilon \F_{mn}\cr
\delta \psi_m &=& 0\cr
\delta \psi &=& -\epsilon \phi
\eea
Indeed these variations close off-shell and they are nilpotent
\bea
\delta^2 &=& 0
\eea
when acting on any of the fields. We will denote the associated supercharge as $Q$. Let us define
\bea
\D_m \Phi &=& \nabla_m \Phi - i [\A_m,\Phi]
\eea
where $\nabla_m$ is a metric connection covariant derivative, so by taking $\Phi$ to be a scalar field with no vector indices $m,n,...$ this derivative becomes
\bea
\D_m \Phi &=& \partial_m \Phi - i [\A_m,\Phi]
\eea
On this adjoint scalar field we define the action of the corresponding  field-strength by
\bea
[\D_m,\D_n] \Phi &=& -i [\F_{mn},\Phi]
\eea
which, in the form notation, becomes 
\bea
\D \wedge \D &=& -i \F
\eea
Explicitly we get
\bea
\F_{mn} &=& F_{mn} - \phi_{mn} - i [\phi_m,\phi_n]\cr
\bar\F_{mn} &=& F_{mn} + \phi_{mn} - i[\phi_m,\phi_n]
\eea 
where we define
\bea
\phi_{mn} &=& D_m\phi_n - D_n\phi_m
\eea
with
\bea
D_m \phi_n = \nabla_m \phi_n -i[A_m, \phi_n]
\eea
The supersymmetric Lagrangian is given by
\bea
\L &=& \L_0 + \L_1
\eea
where
\bea
\L_0 &=& \Tr\bigg[\, \frac{1}{4} \bar\F^{mn} \F_{mn} + \frac{1}{2} \phi^2 - \phi \D^m \phi_m\cr
&&~~~~ + i \psi^m \D_m \psi + i \psi^{mn} \(\bar\D_m \psi_n - \bar\D_n \psi_m\)\bigg]\cr
\L_1 &=& -\Tr\bigg[\frac{i}{2} \epsilon^{mnpqr} \psi_{mn} \D_p \psi_{qr}\bigg]
\eea
The first part of the Lagrangian can be expressed in terms of a fermionic quantity 
\bea
V &=& \Tr\bigg[\frac{1}{2}\bar\F^{mn} \psi_{mn} - \frac{1}{2} \phi \psi - \phi^m \D_m \psi\bigg]
\eea
as
\bea
\L_0 &=& \{Q,V\} 
\eea
The remaining term does not depend on the metric. Hence the stress tensor will be $Q$-exact and the theory will be topological.

The bosonic part of the Lagrangian can be rewritten as
\bea
\L_{boson} &=&   \Tr\bigg[\frac{1}{4} \bar\F^{mn} \F_{mn} - \frac{1}{2} \(\D^m \phi_m\)^2\bigg]
\eea
by integrating out the auxiliary field $\phi$. Furthermore
\bea
\frac{1}{4} \bar\F^{mn} \F_{mn} &=& \frac{1}{4} F^{mn} F_{mn} - \frac{1}{4} [\phi^m,\phi^n][\phi_m,\phi_n] - \frac{1}{4} \phi^{mn} \phi_{mn} - \frac{i}{2} F^{mn} [\phi_m,\phi_n]
\eea
where
\bea
\frac{1}{4}\phi^{mn}\phi_{mn} &=& \frac{1}{2} D^m \phi^n D_m \phi_n
-\frac{1}{2} D_m \phi_n D_n \phi_m
\eea
Also note that 
\bea
\frac{1}{2}\(\D^m \phi_m\)^2 &=& \frac{1}{2}\(D^m \phi_m\)^2
\eea
Now by  integrations by parts and commuting two covariant derivatives, we can write
\bea
D_m \phi_n D^n \phi^m &=& \(D^m\phi_m\)^2 + \phi^m [D_m,D_n] \phi^n
\eea
where we ignore the total derivative terms on the right hand side. The commutator acting on any vector field $\Phi^n$ can be expressed as
\ben
[D_m, D_n]\Phi^n= [\nabla_m, \nabla_n]\Phi^n -i[F_{mn}, \Phi^n]=-R_{mn} \Phi^n -i[F_{mn}, \Phi^n]
\label{commutator}
\een
Thus collecting all the terms, the bosonic  part of Lagrangian becomes
\ben
\L_{boson} &=&  \Tr\bigg[\frac{1}{4}F_{mn}F^{mn} - \frac{1}{2}D_m\phi_n D^m \phi^n -\frac{1}{2}R_{mn}\phi^m  \phi^n - \frac{1}{4}[\phi_m,\phi_n][\phi^m,\phi^n]\bigg]
\label{bosmas}
\een
In flat space this corresponds to the Lagrangian we get by dimensional reduction of ten dimensional YM Lagrangian with $SO(5,5)$ Lorentz symmetry along 5 timelike directions \cite{Hull:2014cxa}.

\section{Five-manifold with one isometry}
We now assume the existence of a unit normalized Killing vector field $v^m$ on $M$, and as before in 6d, we again define trace parts of our fields as
\bea
\varphi &=& v^m \phi_m\cr
\chi &=& v^m \psi_m\cr
\chi_m &=& v^n \psi_{nm}
\eea
and then we separate our fields into traceless and trace parts as follows
\bea
\phi_m &=& \phi'_m + v_m \varphi\cr
\psi_m &=& \psi'_m + v_m \chi\cr
\psi_{mn} &=& \psi'_{mn} + v_m \chi_n - v_n \chi_m
\eea

In presence of a Killing vector we shall require the existence of a $\mb{Z}_2$ symmetry that acts on the fields as
\bea
\psi &\leftrightarrow & \chi\cr
\psi'_m &\leftrightarrow & -2 \chi_m\cr
\psi'_{mn} &\leftrightarrow & -\frac{1}{2} \epsilon_{mnpqr} \psi'^{pq} v^r\cr
\varphi &\leftrightarrow & \varphi\cr
\phi'_m &\leftrightarrow & -\phi'_m
\eea
The $\mb{Z}_2$ leaves $A_m$ unchanged, and so there is no need to separate $A_m$ into traceless and trace parts.

\subsection{The Bosons}
The first and the last term in (\ref{bosmas}) are $\mb{Z}_2$ invariant. The third term in (\ref{bosmas}) is expanded as
\ben
 \Tr\bigg[-\frac{1}{2}R_{mn} \phi^m\phi^m\bigg] =  \Tr\bigg[-\frac{1}{2}R_{mn} {\phi'}^m {\phi'}^n - \frac{1}{2} R_{mn} v^m v^n \varphi^2
- R_{mn} v^m {\phi'}^n \varphi\bigg]
\label{curvnon}
\een 
where only the last term is not $\mb{Z}_2$ invariant. It remains to analyze the second term in (\ref{bosmas}). By expanding this out, we find 
\bea
 \Tr\bigg[-\frac{1}{2}
D_m  {\phi}_n  D^m  {\phi}^n\bigg] &=& \Tr\bigg[-\frac{1}{2}\(
D_m  {\phi'}_n  D^m  {\phi'}^n  +D_m \varphi D^m \varphi 
+ \frac{1}{4}w_{mn}w^{mn}\varphi^2 \)\cr 
&-& w^{mn} D_m \phi'_n \varphi  - \frac{1}{2} (\nabla^m w_{mn})  {\phi'}^n \varphi\bigg]
\eea
where we ignore the total derivative contribution. The terms in the second line are not $\mb{Z}_2$ invariant but
its last term cancels against the $\mb{Z}_2$ non-invariant term in (\ref{curvnon}) upon using the relation (\ref{kcurv}). Therefore we conclude that only the following contribution
\bea
-  \Tr\bigg[w^{mn} D_m \phi'_n \varphi\bigg]
\eea
is not $\mb{Z}_2$ invariant.

\subsection{The Fermions}
We first expand the Lagrangian as
\bea
&& \Tr\bigg[\(i\psi'^m D_m \psi - 2i \chi^m D_m \chi\) + \(2i \psi'^{mn} D_m \psi'_n - 2i \epsilon^{mnpqr} v_m \chi_n D_p\psi'_{qr}\)\cr
&&\,  + \,\(iv^m \chi D_m \psi\) +\(2i v^m \chi^n D_m \psi'_n\) -\(2i v^n \chi^m D_m \psi'_n\) \cr
&&\, -\(\frac{i}{2}\epsilon^{mnpqr} \psi'_{mn} D_p \psi'_{qr}\)\cr
&&\, - \(\psi'^m [\phi'_m,\psi]+2\chi^m [\phi'_m,\chi]\) -\(\chi [\varphi,\psi]\) + \( 2 \chi^n [\varphi,\psi'_n]\)\cr
&&\, +\(2\psi'^{mn} [\phi'_m,\psi'_n]-2\epsilon^{mnpqr} \chi_n[\phi'_p,\psi'_{qr}]v_m\)\cr
&&\, +\(\frac{1}{2}\epsilon^{mnpqr} \psi'_{mn}[\varphi,\psi'_{qr}]v_p\)\cr
&&\, +\,\, i \psi'^{mn} w_{mn} \chi-i \epsilon^{mnpqr} v_m w_{pq} \chi_n \chi_r\bigg]
\eea
and we find that the last line is $\mb{Z}_2$ non-invariant. The term in the third line is $\mb{Z}_2$ invariant but showing that is not  that straightforward. 
To see this, let us first introduce a notation 
\bea
\tilde{\psi}'_{mn} = -\frac{1}{2}\epsilon_{mnabc} {\psi'}^{ab} v^c
\eea
Using this definition, one can show that
\bea
\epsilon^{mnabc} \tilde{\psi}'_{mn} D_a \tilde{\psi}'_{bc} =
\epsilon^{mnabc} {\psi}'_{mn} D_a {\psi}'_{bc} +\epsilon^{mnabc}  {\psi}'_{mp} w^{p}\,_n 
\psi'_{ab} v_c
\eea
The second term then can be written in a $\mb{Z}_2$ invariant form as
\bea
+2 \psi'^{m}\,_p w^{pq}  \tilde{\psi}'_{qm} = \psi'^{m}\,_p w^{pq}  \tilde{\psi}'_{qm}
+\tilde{\psi}'^{m}\,_p w^{pq} {\psi}'_{qm}
\eea
Therefore we can write the third line in a manifestly $\mb{Z}_2$ invariant form as
\bea
\frac{i}{2}\epsilon^{mnpqr} \psi'_{mn} D_p \psi'_{qr}&=&
\frac{i}{4}\epsilon^{mnpqr} \psi'_{mn} D_p \psi'_{qr}+
\frac{i}{4}\epsilon^{mnpqr}\tilde{\psi}'_{mn} D_p \tilde{\psi}'_{qr}\cr
&-& \frac{i}{4}\psi'^{m}\,_p w^{pq}  \tilde{\psi}'_{qm}
\ -\ \frac{i}{4}\tilde{\psi}'^{m}\,_p w^{pq} {\psi}'_{qm}
\eea
Collecting all the non-invariant terms, we have 
\bea
\L^{Bose+Fermi}_{noninvariant} &=&  \Tr\bigg[-\frac{1}{2}w^{mn} \phi'_{mn} \varphi + i w^{mn} \psi'_{mn} \chi - i \epsilon^{mnpqr} v_m w_{pq} \chi_n \chi_r\bigg]
\eea
We can write this in the following form
\bea
\L^{Bose+Fermi}_{noninvariant} &=& -\Delta \L+ \L_{inv}
\eea
with
\ben
\Delta \L = -\left\{Q, \Tr\bigg[w^{mn} \psi_{mn} \varphi +\frac{1}{4} \epsilon^{mnpqr} v_m w_{pq} \psi_n \phi_r\bigg]\right\} \label{non}
\een
where we have gathered all non-invariant terms into a $Q$-exact term, and the rest
\bea
\L_{inv} &=&  \Tr\bigg[-\frac{1}{2} w^{mn} \(F_{mn}-i[\phi_m,\phi_n]\)\varphi 
+\frac{1}{2} w_{mn} w^{mn} \varphi^2
- i \epsilon^{mnpqr} v_m w_{pq} \(\chi_n \chi_r + \frac{1}{4} \psi_n \psi_r\)\bigg]
\eea
is invariant. 

To get a $\mb{Z}_2$ invariant Lagrangian, we add the $Q$-exact term $\Delta \L$ in (\ref{non}) to the Lagrangian that will be also invariant under the $Q$ supersymmetry. But that implies that it will also be invariant under the new supersymmetry that we will denote as $Q^{\vee}$ that is obtained by transforming the fields in the original $Q$ transformation laws by the $\mb{Z}_2$ symmetry.  

For later use, we here write down the full bosonic part of the resulting $\mb{Z}_2$ invariant Lagrangian,
\ben
\L_B &=& \Tr\bigg[\frac{1}{4} F^{mn} F_{mn} - \frac{1}{4} \phi'^{mn} \phi'_{mn} - \frac{1}{2} (D^m \phi'_m)^2 - \frac{i}{2} F^{mn} [\phi'_m,\phi'_n]\cr
&& - \frac{1}{2} D^m \varphi D_m \varphi  
- \frac{1}{4} [\phi'^m,\phi'^n][\phi'_m,\phi'_n] - \frac{1}{2} [\varphi,\phi'^m][\varphi,\phi'_m]\cr
&& + \frac{1}{4} w^{mn} w_{mn} \varphi^2 - \frac{1}{2} w^{mn} \(F_{mn} - i[\phi'_m,\phi'_n]\) \varphi\bigg]\label{LB}
\een
where we define $\phi'_{mn} = D_m \phi'_n - D_n \phi'_m$, and the full $\mb{Z}_2$ invariant fermionic Lagrangian,
\ben
\L_F &=& \Tr\bigg[i \psi^m \D_m \psi + 2i \psi^{mn} \bar\D_m \psi_n\cr
&& - \frac{i}{2} \epsilon^{mnpqr} \psi_{mn} \D_p \psi_{qr}\cr
&& - i w^{mn} \psi_{mn} \chi - \frac{i}{4} \epsilon^{mnpqr} v_m w_{pq} \psi_n \psi_r\bigg]\label{LF}
\een

Let us notice that the bosonic part of the Lagrangian can be recast in the following form
\ben
\L_B &=& \Tr\bigg[\frac{1}{4} \t F^{mn} \t F_{mn}\cr
&& - \frac{1}{4} \phi'^{mn} \phi'_{mn} - \frac{1}{2} (D^m \phi'_m)^2\cr
&& - \frac{1}{2} D^m \varphi D_m \varphi - \frac{1}{2} [\varphi,\phi'^m][\varphi,\phi'_m]\bigg]\label{LB5d}
\een
where we introduce a background two-form $B_{mn} = -w_{mn} \varphi - i [\phi'_m,\phi'_n]$ and the gauge covariant field strength
\bea
\t F_{mn} &=& F_{mn} + B_{mn}
\eea
In this form, we can understand this Lagrangian, at least for abelian gauge group, as the dimensional reduction of the 6d Lagrangian by putting time derivatives to zero. More precisely, the tensor piece (\ref{BMN}) of the 6d Lagrangian should reduce to the YM term $\frac{1}{4} \t F^{mn} \t F_{mn}$ (although this we can not directly see because our 6d Lagrangian involves a nonchiral tensor field, but is a consequence of supersymmetry), and then the rest of the 6d Lagrangian can be reduced by simply putting time derivatives to zero on the fields. This dimensional reduction gives us
\bea
\L_B &=& \frac{1}{4} \t F^{mn} \t F_{mn} -\frac{1}{4}\phi_{mn}^2 - \frac{1}{2} (\nabla^m \phi_m)^2 - \frac{1}{4} w_{mn}^2 \varphi^2 + \frac{1}{2} w^{mn} \phi_{mn} \varphi
\eea
that, we can show, agrees with (\ref{LB5d}) for abelian gauge group.

Let us finally notice that the 5d Lagrangian does not agree with the 6d Hamiltonian (\ref{6dH}) when we put time derivatives to zero there\footnote{The 
detailed
reasoning of  comparing the 5d Lagrangian to the 6d Hamiltonian putting  time derivative terms to zero will be explained at the beginning of Section \ref{sec8}.}. 
We will address this problem in the rest of this paper, but we will not be able to solve this problem in Lorentzian signature. We will solve the problem only in Euclidean signature, or for Wick rotated scalar fields in 5d.

\section{The supersymmetry algebra}
In this section, we would like to work out the supersymmetry algebra with the two supersymmetries. Let us begin by decomposing  the first supersymmetry variation as
\bea
\delta \phi'_m &=& -i\psi'_m\cr
\delta \varphi &=& -i \chi\cr
\delta A_m &=& -i \psi'_m - i v_m \chi\cr
\delta \phi &=& 0\cr
\delta \chi_m &=& \frac{1}{2} v^n \F_{nm}\cr
\delta \psi'_{mn} &=& \frac{1}{2} \F'_{mn}\cr
\delta \psi'_m &=& 0\cr
\delta \chi &=& 0\cr
\delta \psi &=& -\phi
\eea
where $\phi = D^m \phi_m$. The corresponding conserved supercurrent is 
\bea
j^m &=& \Tr\bigg[-\F^{mn} \(\psi'_n +  v_n \chi\) + \phi \(\psi'^m + v^m \chi\) + \frac{1}{2} \epsilon^{mabcd} \F_{ab} 
\(\psi'_{cd} + 2 v_c \chi_d\)\bigg]
\eea
whose form is not affected by the presence of the additional correction term (\ref{non}) in the Lagrangian. 
Since the improved Lagrangian is $\mathbb{Z}_2$ invariant, one can map 
these variations by the $\mb{Z}_2$ transformation into a second supersymmetry variation
\bea
\delta^{\vee} \phi'_m &=& -2i \chi_m\cr
\delta^{\vee} \varphi &=& -i\psi\cr
\delta^{\vee} A_m &=& 2i\chi_m - i v_m \psi\cr
\delta^{\vee} \phi^\vee &=& 0\cr
\delta^{\vee} \chi_m &=& 0\cr
\delta^{\vee} \psi'_{mn} &=& -\frac{1}{4} \epsilon_{mnabc} \F'^{\vee ab} v^c\cr
\delta^{\vee} \psi'_m &=& -v^n \F_{nm}^{\vee}\cr
\delta^{\vee} \chi &=& -\phi^{\vee} \cr
\delta^{\vee} \psi &=& 0
\eea
Here we define $\phi^{\vee m} = -\phi'^m + v^m \varphi$ and $\phi^{\vee} = D^m \phi^{\vee}_m$. Utilizing the $\mathbb{Z}_2$ transformation further,
the corresponding supercurrent can be identified as 
\bea
j^{\vee m} &=&  \Tr\bigg[-\F^{\vee mn} \(-2\chi_n + v_n \psi\)+ \phi^{\vee} \(-2 \chi^m + v^m \psi\) - 
\F^{\vee}_{ab} \(\psi'^{ab} v^m + 2 \psi'^{ma} v^b\)\cr
&& + \frac{1}{2} \epsilon^{mabcd} \F^{\vee}_{ab}\,  \psi'_c \, v_d\bigg]
\eea

\subsection{Anti-commutator of two supercharges }
Since we know that $\delta^2 =0$ formally\footnote{There is an extra  central term contribution, which will be identified shortly.}, one has
$\delta^{\vee 2}=0$ as it should be due to the $\mathbb{Z}_2$ symmetry of our Lagrangian. Below we shall find that $\{\, \delta, \delta^\vee \}$ 
acting on the fields generates a translation along the Killing vector direction $v^m$ up to some gauge transformation. Since our fields are in general 
tensors in the five space, the translation is in general generated by the Lie derivative ${\cal L}_v$ along the $v^m$ direction. In showing this algebra, we find that one needs
equations of motion for some cases. Since the equations of motions are affected by the improved terms, this computation will justify the correctness
of the improved term to some extent.  Below we shall illustrate the computation of $\{\, \delta, \delta^\vee \}$ focusing on the cases where
 the equations of motion are necessary.

Let us record here the equations of motion first: They reads
\ben
 \D_m \bar\F^{mn}  +\bar\D^n \phi + 2\{\psi^n, \psi \} -  \epsilon^{nabcd} \{\psi_{ab},\psi_{cd}\}-\frac{v^n}{2} \F_{pq} w^{pq}
-2 \D_m(w^{mn} \varphi) &=& 0
\label{b1}
\\
 \bar\D_m \F^{mn} -\D^n \phi +4 \{\psi_m,\psi^{nm}\}
+\frac{v^n}{2} \F_{pq} w^{pq}
 &=& 0
\label{b2}
\\
\phi - \D^m \phi_m &=& 0
\label{b3}
\een
and the fermionic are
\ben
\D_n \psi + 2 \bar\D^m \psi_{mn}+ v_n \, w^{pq}\psi_{pq}+ \frac{1}{2}\psi^m \epsilon_{mnabc} w^{ab} v^c   &=& 0 \label{f1} \\
\bar\D_m \psi_n - \bar\D_n \psi_m - \epsilon_{mnpqr} \D^p \psi^{qr} -w_{mn} \chi &=& 0  \label{f2}\\ 
\D_m \psi^m &=& 0  \label{f3}
\een

Let us begin with the case of $\psi$: Using the rules of supersymmetry transformation, one finds
\bea
\{\, \delta, \delta^\vee \} \psi = i v^m \D_m \psi  + 2 i \bar\D_m \chi^{m}
\eea
Now we contract $v^n$ on Eq.~(\ref{f1}) and obtain 
\bea
v^n \D_n \psi = 2 \bar\D_m \chi^m
\eea 
It should be noted that, in this computation, the third term of (\ref{f1}) is essential, which is from the improved term required by the
$\mathbb{Z}_2$ symmetry. 
Using this equation, one finds
\bea
\{\, \delta, \delta^\vee \} \psi =2 i v^m \D_m \psi  =2 i (\L_v \psi -i[\Lambda , \psi] ) 
\eea
where  the gauge parameter  is given by $\Lambda = a -\varphi$ with $a= v^m A_m$.
Hence we conclude that the anticommutator of the two supercharges
generates the translation together with the gauge transformation by a gauge parameter $\Lambda$.  In addition, 
one can say that the improved term plays the role which makes the equation of motion
in the right form for the  generation of  the translation.  For the case of $\chi$, again one needs the equations of motion: By a straightforward
computation, one finds
\bea
 \{\, \delta, \delta^\vee \} \chi = i v^m \D_m \chi  - i \bar\D_m \psi'^{m} =2 i (\L_v \chi -i[\Lambda , \chi] ) 
\eea
where we used (\ref{f3}) for the second equality. The third  is for the case of  $\psi'_{mn}$: Without using the equation of motion, 
one finds
\bea
\{\, \delta, \delta^\vee \} \psi'_{mn} = 2i \left(
(\D_m \chi_n -\D_n \chi_m )' + \frac{1}{2} \epsilon_{mnabc} \bar\D^a \psi'^b \, v^c
\right)
\eea 
where the extra prime denotes a transverse part  of the tensor inside the parenthesis. Using (\ref{f2}),
one finds
\bea
\{\, \delta, \delta^\vee \} \psi'_{mn}= 2 i (\L_v \psi'_{mn} -i[\Lambda , \psi'_{mn}] ) 
\eea
where again the improved term plays an essential role for this to generate the translation. For the rest of the fields, the equations 
of motion are not 
necessary to show the above: Namely one can show that
\bea
\{\, \delta, \delta^\vee \} a= 2 i (\L_v \, a -v^m D_m \Lambda ) 
\eea
and
\bea
\{\, \delta, \delta^\vee \} A'_m= 2 i (\L_v \, A'_m - D'_m \Lambda ) 
\eea
for the case of the gauge fields. For the further remaining non gauge fields which we collectively denote by $\Phi$ , one finds
\bea
\{\, \delta, \delta^\vee \} \Phi= 2 i (\L_v \, \Phi -i[ \Lambda, \Phi] ) 
\eea
without using any equations of motion.

\section{BPS equations and 4D instantons}

We can obtain BPS equations by putting fermions to zero and a certain linear combination of the supersymmetry variations of fermions to zero. That is, by putting 
\bea
(\delta + \alpha \delta^{\vee}) \psi_{mn} &=& 0\cr
(\delta + \alpha \delta^{\vee}) \psi_m &=& 0\cr
(\delta + \alpha \delta^{\vee}) \psi &=& 0
\eea
Thus the BPS equations will depend on one parameter $\alpha$. These BPS equations should be supplemented by the 6d Gauss law which in 5d is the equation of motion for $A_m$.

Let us first show that the BPS equations together with the 6d Gauss law constraint imply that the full equations of motion are obeyed automatically. 
The 6d Gauss law constraint refers to the equation of motion following from adding (\ref{b1}) and (\ref{b2}), and reads explicitly 
\ben
D^m \(F_{mn} - w_{mn} \varphi\) &=& i [\phi'^m,D_n \phi'_m] + i [\varphi,D_n\varphi]\label{6dGauss}
\een
and we shall refer to its trace part 
\ben
v^n D^m \(F_{mn} - w_{mn} \varphi\) &=& i [\phi'^m, v^n D_n \phi'_m] + i [\varphi, v^n D_n\varphi]
\label{5dGauss}
\een
as 5d Gauss law constraint.
Let us begin with the  $\alpha=0$ case. The BPS equations read 
\bea
D^m\phi_m & =& 0
\cr
\F_{mn}&= &0
\eea
while setting all the fermion fields to zero. Then one finds that (\ref{b2}) is trivially satisfied with the BPS equations. Thus with the 6d Gauss law constraints the full equations of motions are obeyed.

For nonzero values of $\alpha$, the BPS equations read
\ben
&& D^m\phi_m = D^m \phi^{\vee}_m=0
\cr
&& v^n \F_{nm}= v^n \F^{\vee}_{nm}=0
\een
as well as 
\ben
\F'_{mn}= \frac{\alpha}{2} \epsilon_{mnpqr} \F^{\vee pq}\, v^r
\label{bpsb1}
\een
while setting all the fermion fields to zero.  First we note that
\ben
\bar{\D}_m{\F}^{mn}=\bar{\D}_m{\F'}^{mn} = \frac{\alpha}{4} \epsilon^{mnpqr} \F^{\vee}_{pq}\, w_{mr}
\label{alpha1}
\een 
where we used the fact ${\F}^{mn}={\F'}^{mn}$ on the BPS equations and
\bea
 \epsilon^{mnpqr}(\bar{\D}_m \F^{\vee}_{pq})\, v_{r}=0
\eea
This last identity follows from the fact 
\ben
\F^{\vee}_{mn}=\bar\F_{mn}-2\bar{\D}_m(v_n \varphi)+2\bar{\D}_n(v_m \varphi)\label{veebar}
\een
together with the Bianchi identity. The BPS equation (\ref{bpsb1}) can also be presented as
\bea
{\F'}^\vee_{mn}={\F}^\vee_{mn}= \frac{1}{2\alpha} \epsilon_{mnpqr} \F^{pq}\, v^r
\eea
Inserting this into the right side of (\ref{alpha1}), we find
\bea
\bar{\D}_m{\F}^{mn}= - \frac{v^n}{2} w_{pq}\F^{pq}
\eea
which agrees with  (\ref{b2}) with $\phi=0$. For this case, it is straightforward to show that
the 5d Gauss law constraint is reduced to
\ben
w^{mn}\( \F_{mn}+\bar{\F}_{mn} -2 w_{mn}\varphi\)=4 \D^m \bar\D_m \varphi
\label{6dgauss}
\een 
while the traceless part of the 6d Gauss law constraint is automatically satisfied once the BPS equations are obeyed.  

The $\alpha=\infty$ case can be treated by the $\mathbb{Z}_2$ transformed version of 
$\alpha=0$ case leading to the same conclusion.

\subsection{4D instantons}
In this section we would like to study BPS equations describing 4D instantons
of the $\mathbb{Z}_2$ invariant system. To be specific, we shall assume
the metric 
\bea
ds^2 &=& (dx^5 + v_i dx^i)^2 + g_{ij} dx^i dx^j
\eea
where $v_i$ and the four metric $g_{ij}$ are independent of $x^5$ coordinates. The Killing vector is then explicitly given by
$v^m =(0,0,0,0,1)$ or by $v_m=(v_i, 1)$.
We have the metric component
$G_{55} =1$, $G_{i5}=v_i$
and the inverse metric has components
\bea
G^{55} &=& 1 + g^{ij} v_i v_j\cr
G^{i5} &=& -g^{ij} v_j\cr
G^{ij} &=& g^{ij}
\eea
together with $\sqrt{G}=\sqrt{g}$.
The (anti) instanton BPS equations can be obtained by setting
$(Q\pm Q^\vee)$ variation of the fermionic fields to zero, from which  one finds
\ben
&& D^m\phi_m = D^m \phi^{\vee}_m=0
\cr
&& v^n \F_{mn}=v^n \F^{\vee}_{mn}=0
\label{bpsa}
\een
as well as 
\ben
\F'_{mn}= \pm\frac{1}{2} \epsilon_{mnpqr} \F'^{\vee pq}\, v^r
\label{bpsb}
\een
which should be supplemented by the reduced 5d Gauss law constraint (\ref{6dgauss}). 

Assuming $\phi_m= A_m v^m=0$ and $\partial_5=0$, 
the above set of BPS equations are reduced to the usual 4d (anti) self-dual
equation
\bea
F_{ij} =\pm \frac{1}{2} \epsilon_{ij}\,^{kl} \,  F_{kl} 
\eea
where indices are raised or lowered by 4d metric $g_{ij}$.
But compared to the usual case of 4d instantons, we have an additional constraint (\ref{6dgauss})
\ben
w_{ij} \, F^{ij}=0
\label{wij}
\een
which comes from the 5d Gauss law constraint (\ref{5dGauss}) together with the BPS equations. Hence the usual 4d (anti) instanton 
solutions cannot be solutions of our system unless this additional requirement holds. 

Especially considering the case of $S^5$, one has
\bea
w_{mn}=\frac{1}{2} \epsilon_{mnabc}w^{ab}v^c
\eea
or, in the 4d notation, $w_{ij}$ is self-dual as
\bea
w_{ij} = \frac{1}{2} \epsilon_{ij}\,^{kl} \,  w_{kl} 
\eea
Hence self-dual instanton solutions are not favored due to this extra constraint whereas  the anti self-dual ones are completely free of
it.  
Further study is required for the detailed structure of nature of instantons.

The discussion so far is a bit unsatisfactory since we have not identified the Hamiltonian and its BPS bound, or identified the BPS solutions as saddle points of the Euclidean Lagrangian. Now the Lagrangian is indefinite if we take $\phi_m$ real, but this can be cured by Wick rotating the integration contour to the imaginary axis. But in that case we get an imaginary term in the Lagrangian 
\ben
\L_{w} &=& \frac{i}{2} \Tr\bigg[w^{mn} \(F_{mn} + i [\phi'_m,\phi'_n]\)\varphi\bigg]\label{w}
\een
This is however reflection positive: Under time reversal we have $F_{mn} \rightarrow -F_{mn}$ and $i[\phi'_m,\phi'_n]\rightarrow i[\phi'_m,\phi'_n]$ and by a subsequent complex conjugation we see that the combination $i\(F_{mn} + i [\phi'_m,\phi'_n]\)$ is invariant. But since the term is imaginary, it is not clear what it shall mean to find saddle points of the Lagrangian. We also have not shown that these (selfdual or antiselfdual) instantons are the most general BPS solutions that carry nonzero instanton number. We will address and answer all these questions below.

\section{Time along $v^m$ and corresponding charges}

We will now assume that Euclidean time is along the Killing vector $v^m$. This enable us to derive conserved charges directly in the 5d theory. In particular there will be a Hamiltonian that generates translation along $v^m$. This Hamiltonian can be obtained in two ways. One way is by making supersymmetry variations of the supercurrents. The other way is to use the Noether
method.  Below we shall find that the two results agree with each other as required by the consistency of our formulation. Since 
the computation is based on the fairly standard field-theoretic methods starting from the Lagrangian of the system, we shall only sketch 
its procedure and omit details of derivation.
 
First we obtain the following supersymmetry variations of the supercurrents,
\ben
\delta \, j^m \,\, &=&  \Tr\bigg[\frac{1}{4} \epsilon^{mabcd} \F_{ab} \F_{cd}\bigg] \label{algebra1}\\
\delta^\vee j^{\vee m} &=&  \Tr\bigg[\frac{1}{4} \epsilon^{mabcd} \F^{\vee}_{ab} \F^{\vee}_{cd}\bigg]\label{algebra2}\\
\delta j^{\vee m} &=&  \Tr\bigg[-\frac{1}{2} v^m \F^{\vee}_{ab} \F'^{ab} + v^p \F^{\vee mn} \F_{pn} + v^p \F'^{mn} \F^{\vee}_{pn}
 \cr
&& + \(\F^{\vee mn}\phi + \F^{mn}\phi^{\vee}\) v_n - v^m \phi^{\vee} \phi\bigg]+2 J^m_{F}  \label{algebra3}
\\
\delta^\vee j^m &=&  \Tr\bigg[-\frac{1}{2} v^m \F_{ab} \F'^{\vee ab} + v^p \F^{mn} \F^{\vee}_{pn} + v^p \F'^{\vee mn} \F_{pn}\cr
&& +  \(\F^{mn} \phi^{\vee} + \F^{\vee mn}\phi\) v_n - v^m \phi \, \phi^{\vee}\bigg] +2 J^{\vee m}_{F}
\label{algebra4}
\een
where $ J^m_{F} $ and $J^{\vee m}_{F}$ are the contributions involving fermion fields.
By writing 
\bea
\F'_{mn} = \F_{mn} - v_m \F_n + v_n \F_m
\eea
and by using cyclicity of trace, we see that 
\bea
\delta\,  j^{\vee m} = \delta^\vee j^m &=&  \Tr\bigg[- \frac{1}{2} v^m \F_{ab} \F^{\vee ab} + v^p \F^{mn} \F^{\vee}_{pn} + v^p \F^{\vee mn} \F_{pn}\cr
&+&  \(\F^{mn}\phi^{\vee} + \F^{\vee mn}\phi\) v_n - v^m \phi^{\vee} \phi\bigg] + 2 J^m_{F}
\eea
where the equality works for the purely bosonic contributions as well as for those involving fermionic fields. From this we identify the Hamiltonian current $J^m$  as
\bea
J^m &=&  \Tr\bigg[\frac{1}{2}v^p \F^{mn} \F^{\vee}_{pn} + \frac{1}{2} v^p \F^{\vee mn} \F_{pn} + 
\frac{1}{2} \(\F^{mn}\phi^{\vee} + \F^{\vee mn}\phi\) v_n\cr
&-& v^m \left( \frac{1}{4}  \F_{ab} \F^{\vee ab} + \frac{1}{2} \phi^{\vee} \phi \right)\bigg]\ + \ J^m_{F}
\eea 
which can be shown to be conserved on shell. 

The second method is based on the Noether procedure. We first note that our $\mathbb{Z}_2$ invariant Lagrangian possesses the translation symmetry defined by
$\delta_v \Phi = \L_v \Phi + \delta_\Lambda \Phi$
where $\delta_\Lambda$ denotes the gauge transformation by the gauge parameter $\Lambda= a-\varphi$. Then the Hamiltonian current 
$J^m$ can be precisely reproduced by using Noether's procedure.

Now let $x^5$ as a coordinate along the Killing vector direction defined by
\ben
\frac{\partial}{\partial x^5} = v\label{Killingvector}
\een
and the remaining coordinates along the base four manifold as $x^i\, (i=1,2,3,4)$. Then from (\ref{algebra3}) and (\ref{algebra4}),
one can read off the supersymmetry algebra as
\bea
\{Q,\, Q^\vee \} = - 2 H
\eea
where the Hamiltonian is given by 
\bea
H &=& -\int_{M_4} d^4 x \sqrt{g} \, J^5
\eea 
where $g = G$ is the metric on the five-manifold with a Killing vector (\ref{Killingvector}), which thus is equal to the metric on the contact plane. This is conserved (i.e. independent of  $x^5$ coordinate), which is ensured by the covariant conservation of the current $\nabla_m J^m=0$ when the contact plane $M_4$ is compact, or more generally when the boundary term $\int_{M_4} d^4 x\partial_i (\sqrt{g}J^i)$ is zero. 

Now let us turn to the supersymmetry algebra related to (\ref{algebra1}) and (\ref{algebra2}); Both currents appearing in the right hand side are identically conserved without resorting to any equations of motion. Then the corresponding supersymmetry algebra is identified as
\bea
\{\, Q,\, Q\} = 2Z, \ \ \ \ \ \
\{ Q^\vee, Q^\vee \} = 2 Z^\vee
\eea
with central charges 
\bea
Z\,\, &=& \frac{1}{4}\int_{M_4} d^4 x  \, \epsilon^{ijkl} \Tr\bigg[\F_{ij} \F_{kl}\bigg] \cr
Z^\vee &=& \frac{1}{4}\int_{M_4} d^4 x  \, \epsilon^{ijkl} \Tr\bigg[\F^\vee_{ij} \F^\vee_{kl}\bigg]
\eea
Below we shall verify that
\ben
Z=Z^\vee= \frac{1}{4}\int_{M_4} d^4 x  \, \epsilon^{ijkl} \Tr\bigg[F_{ij} F_{kl}\bigg]\label{BPSbound}
\een
when $M_4$ is closed. This central charge is then counting  the instanton number of  4d configurations.  In more detail, let us first note
\bea
\epsilon^{ijkl} \Tr\bigg[\F_{ij} \F_{kl}\bigg] &=& \epsilon^{ijkl} \Tr\bigg[\(F_{ij} - 2 D_i \phi_j - i [\phi_i,\phi_j]\) \(F_{kl} - 2 D_k \phi_l - i [\phi_k,\phi_l]\)\bigg]\cr
&=& \epsilon^{ijkl} \Tr\bigg[F_{ij} F_{kl}\bigg] + \nabla_i K^i 
\eea
where 
\bea
K^i &=& 4 \epsilon^{ijkl} \Tr \bigg[\phi_j D_k \phi_l +\frac{2i}{3} \phi_k \phi_j \phi_l -\phi_j F_{kl}
\bigg] 
\eea
The integral of the total derivative term $\nabla_i K^i$ can be non-zero on a noncompact five-manifold. This was analyzed in \cite{Lee:2006gqa}. 
Since $K^i$ is gauge invariant and globally defined, on a compact five-manifold the total derivative term can not give any non-vanishing contribution, and we just have the pure instanton number. In a similar manner, one has 
\bea
\epsilon^{ijkl} \F^{\vee}_{ij} \F^{\vee}_{kl} &=&  \epsilon^{ijkl} \(F_{ij} - 2 D_i \phi^\vee_j - i [\phi^\vee_i,\phi^\vee_j]\) 
\(F_{kl} - 2 D_k \phi^\vee_l - i [\phi^\vee_k,\phi^\vee_l]\)\cr
&=& \epsilon^{ijkl} F_{ij} F_{kl} + \nabla_i K^{\vee i}
\eea
where again  $K^{\vee i}$ is a globally defined gauge invariant quantity which may be obtained by replacing $\phi_k$ dependence of $K^i$ by  
$\phi^\vee_k$. Hence the central charge in this case simply counts the instanton number associated with  the gauge bundle of $M_4$.  
 
\subsection{A comment on the Majorana condition} 
We see that if $Q$ is hermitian, then $Q^2\geq 0$ which would imply that the instanton number is positive definite. This seems strange, and we will now show that the supersymmetry algebra itself is inconsistent if we impose the wrong Majorana condition of the supercharges to be associated with having $x^5$ as time direction. 

Let us first consider the usual supersymmetry algebra in $1+1$ dimensions with metric $g_{00} = -1, g_{11} = 1, g_{01} = 0$ and with gamma matrices $\gamma^0 = i \sigma^2$, $\gamma^1 = \sigma^1$ and charge conjugation matrix $C = \gamma^0$ in the Majorana representation. If we impose Majorana condition on the supercharges, $Q_{\pm}^{\dag} = Q_{\pm}$, then we have the supersymmetry algebra
\bea
\{Q_{\pm},Q_{\pm}\} &=& 2 \(H \mp P\)
\eea
where by positivity of the left-hand side, we see that $H$ is the positive semidefinite Hamitonian generating translations along $x^0$ and $P$ is the indefinite momentum generating translations along $x^1$ and we have the BPS bound $H\geq |P|$. We can realize this supersymmetry algebra explicitly by a $(1,0)$ sigma model, in which case we have 
\bea
Q_{\pm} &=& i \int dx^1 \psi_{\pm}^i \(\partial_0 \pm \partial_1\) X^i\cr
H &=& \int dx^1 \frac{1}{2} \((\partial_0 X^i)^2 + (\partial_1 X^i)^2\)\cr
P &=& \int dx^1 \partial_0 X^i \partial_1 X^i
\eea

If we instead take $x^1$ as time direction in this sigma model, and use the same Majorana representation as before, and assume that supercharges are Majorana as before, then we find the supersymmetry algebra
\bea
\{Q_{\pm},Q_{\pm}\} &=& 2 \(P \mp H\)
\eea
where
\bea
Q_{\pm} &=& i \int dx^0 \psi_{\pm}^i \(\partial_1 \pm \partial_0\) X^i\cr
H &=& \int dx^0 \frac{1}{2} \((\partial_1 X^i)^2 + (\partial_0 X^i)^2\)\cr
P &=& \int dx^0 \partial_1 X^i \partial_0 X^i
\eea
Now the supersymmetry algebra is inconsistent with positivity of the left-hand side since $P$ can be indefinite. Let us now define
\bea
Q_{\pm} &=& \frac{1}{\sqrt{2}} \(Q \pm Q^{\vee}\)
\eea
Then the supersymmetry algebra reads
\bea
\{Q,Q^{\vee}\} &=& -2H\cr
\{Q,Q\} &=& 2Z\cr
\{Q^{\vee},Q^{\vee}\} &=& 2Z
\eea
which is exactly the algebra that we found. 

One way to resolve the contradiction is to also replace the Majorana condition with 
\bea
\psi^{\dag} \gamma^1 &=& \psi^T \gamma^0
\eea
We then get
\bea
Q_{\pm}^{\dag} &=& \mp Q_{\pm}
\eea
and the supersymmetry algebra becomes consistent with positivity,
\bea
\{Q_{\pm},Q^{\dag}_{\pm}\} &=& 2\(H\pm P\)
\eea
Translated into the supercharges $Q$ and $Q^{\vee}$, we should impose the reality conditions
\bea
Q^{\dag} &=& -Q^{\vee}\cr
Q^{\vee \dag} &=& -Q
\eea
This discussion applies to the case when time and space directions are orthogonal. This is not the case if $G_{5i} = v_i$ is nonvanishing, which  
we are mainly interested in here. In this case we find additional terms in the Hamiltonian that are indefinite roughly 
corresponding to momentum. There is no clean separation of this indefinite part in the Hamiltonian from the positive semidefinite part in Lorentzian signature. But if we Wick rotate the scalar fields, then the indefinite part becomes purely imaginary and we have a clean separation. In the next section, we will see another stronger argument why we shall apply this Wick rotation.

Since we do not have the expected Majorana condition in this approach, we will not attempt to pursue this direction further. What we have achieved so far is that we have identified the BPS bound as the instanton charge. Due to its topological nature, this BPS bound will be the same regardless we pick $x^5$ or $x^0$ as our time direction with respect to which we define our charges. We will also see this below by explicit computations.

\section{Time along $x^0$ and corresponding charges}\label{sec8}
In 5d the 6d Gauss law (\ref{6dGauss}) is nothing but the $A_m$ equation of motion. In 5d we can consider the potential energy of the 6d Hamiltonian. If the fields are momentarily static, then the potential energy will be equal to the full Hamiltonian. At the next instant of time the fields will start to evolve with time if the initial field configuration was not at a minimum of the potential energy, and this time evolution is difficult to capture directly in the 5d theory. But we can at least obtain the potential energy of a static field configuration. 

We can not use the potential energy to derive the Hamilton equations of motion and we can not use it for quantizing the theory. Another indication that we can not directly use the potential energy as a Hamiltonian to quantize the theory, comes from the Gauss law. In 5d this is nothing but the full equations of motion for the gauge potential $A_m$. This is certainly problematic, since if we use the equation of motion for $A_m$ in a path integral quantization, then we are really just considering the classical field configuration, and we do not include any quantum fluctuations there, which surely is an incorrect way of quantizing the 5d SYM theory. 

Still we can consider the potential energy as a classical  field configurations that satisfy the classical 5d equations of motion, and in particular the $A_m$ equation of motion which is the 6d Gauss law. Here it will be useful and have a physical significance as the potential energy of the system, and it can be used to understand the BPS bound and to derive BPS equations, since these are classical equations of motion. 
 
The potential energy should be equal to the 5d Lagrangian upon dimensional reduction along time. For our case of 5d SYM, we will encounter a problem since our Lagrangian is not positive semidefinite. But the energy in a supersymmetric theory must be bounded from below by zero. The only resolution seems to be that we Wick rotate $\phi_m$ to the imaginary axis, and our claim is that this Wick rotated Lagrangian is the 6d potential energy of the corresponding classical field configuration. It is also important to note that the Hamiltonian is an operator that does not see the signature of the time direction, so the potential energy will be the same in both Euclidean and Lorentzian signature.

We thus Wick rotate $\phi_m$ to the imaginary axis, and for notational convenience, let us do it in such a way that
\bea
\A_m &=& A_m + i \phi_m
\eea
for the new rotated hermitian field $\phi_m$. Then we find that the 6d 
Gauss law is more restrictive and this in turn cures the problem we found above. The Gauss law (\ref{6dGauss}) splits into two separate conditions
\bea
D^m F_{mn} &=& -i [\phi'^m,D_n \phi'_m] - i [\varphi,D_n \varphi]\cr
D^m \(w_{mn} \varphi\) &=& 0
\eea
We expand out the second condition  
\bea
(D^m w_{mn}) \varphi + w_{mn} D^m \varphi &=& 0
\eea
If we contract this by $v^n$, then we get
\bea
w_{mn} w^{mn} \varphi &=& 0
\eea
Note that the same conclusion follows from the 5d version of the Gauss law constraint (\ref{5dGauss}).
If $w_{mn} = 0$, then we have a trivial line bundle and the situation is essentially that of \cite{Kapustin:2006pk}. In particular, the correction term (\ref{w}) will be absent. Let us therefore assume that $w_{mn} \neq 0$. In this case we get 
\bea
\varphi &=& 0\cr
{\rm Re}(\F_{mn}) &=& F_{mn} + i [\phi'_m,\phi'_n]\cr
{\rm Im}(\F_{mn}) &=& D_m \phi'_n - D_n \phi'_m
\eea
The Hamiltonian is given by the Lagrangian where we put $\varphi = 0$ using the Gauss law,
\ben
H &=& \Tr\bigg[\frac{1}{4} (F_{mn} + i[\phi'_m,\phi'_n])^2 + \frac{1}{4} (D_m \phi'_n - D_n \phi'_m)^2 + \frac{1}{2} (D^m \phi'_m)^2\cr
&=& \frac{1}{4} (F_{mn})^2 + \frac{1}{2} (D_m \phi'_n)^2  + \frac{1}{2}  R^{mn}\phi'_m \phi'_n+ \frac{1}{4} (i[\phi'_m,\phi'_n])^2\bigg]\label{H}
\een
This is manifestly $\mb{Z}_2$ invariant. Since $\varphi = 0$, we also have to satisfy the equation of motion of $\varphi$, which amounts to the additional constraint
\ben
w^{mn} \(F_{mn} + i [\phi'_m,\phi'_n]\) &=& 0\label{added}
\een
For example, if we consider $S^5$ and we pick coordinates so that $w_{mn}$ is a selfdual Kahler form on $\mb{CP}_2$, defined as the contact plane associated with the Killing vector $v_m$, then we find that selfdual instanton $F_{mn} \sim w_{mn}$ is killed by (\ref{added}), although it is difficult to see what implications this condition has on other selfdual instanton solutions.

\section{Vanishing theorem}
To be able to derive a vanishing theorem from the BPS bound, we need to establish that the Hamiltonian can be obtained from a supercharge. We claim that our supercharges, when Euclidean time is along $x^0$, is given by the integral of  
\bea
j^0 &=& \frac{1}{2}\Tr\bigg[\psi (\delta \psi)^{\dag} + \psi_m (\delta \psi^m)^{\dag} + 2 \psi_{mn} (\delta \psi^{mn})^{\dag}\bigg]\cr
j^{\vee 0} &=& \frac{1}{2}\Tr\bigg[\psi (\delta^{\vee} \psi)^{\dag} + \psi_m (\delta^{\vee} \psi^m)^{\dag} + 2 \psi_{mn} (\delta^{\vee} \psi^{mn})^{\dag}\bigg]
\eea
over the five-manifold. There is no direct way that we can prove this since we do not have access to the $x^0$ direction in the 5d theory. We can guess this form by looking at the abelian theory. We can also check that if we make supersymmetry variations of these charges, we reproduce the Lagrangian, that as we have argued is equal to the potential energy, when we use the Euclidean Gauss law that puts $\varphi = 0$. 

First we find by making supersymmetry variations of the above $j^0$ and $j^{0\vee}$, that
\bea
H &=& \int d^5 x \sqrt{G} \, \Tr\bigg[\frac{1}{4} \bar\F^{mn} \F_{mn} + (D^m \phi_m)^2\bigg]\cr
H^{\vee} &=& \int d^5 x \sqrt{G}\,  \Tr\bigg[\frac{1}{4} \bar\F^{\vee mn} \F^{\vee}_{mn} + (D^m \phi^{\vee}_m)^2\bigg]\cr
P &=& \frac{1}{4}\int d^5 x \sqrt{G}\,  \epsilon^{mnabc} \, \Tr\bigg[\bar\F^{\vee}_{mn} \F_{ab}\bigg]v_c
\eea
where $H^{\vee} = (Q^{\vee})^2$ becomes equal to $H = Q^2$ by using the Gauss law that puts $\varphi = 0$, and then it also agrees with (\ref{H}). In particular we see from (\ref{veebar}) that $\bar\F_{mn} = \F^{\vee}_{mn}$ when $\varphi = 0$. It then also follows that 
\bea
P &=& \frac{1}{4} \int d^5 x \sqrt{G}  \epsilon^{mnabc} \Tr\bigg[\F_{mn} \F_{ab}\bigg]v_c
\eea

We find that  
\bea
B(t) &=& H - f(t) P
\eea
is positive semidefinite, where
\bea
f(t)=
\frac{t^{-1} - t}{t^{-1} + t}
\eea
This is a consequence of $(aQ+bQ^{\vee})^2$ being positive semidefinite. Here we restrict ourselves to $t$ real, or $a$ and $b$ real. 
Since $f(t)= f(-t)$, it suffices to consider the $t \ge 0$ branch.

Since $H$ is independent of $t$ we learn that 
\bea
B(u) - B(t) &=& (f(t) - f(u)) P
\eea
If the BPS equations are satisfied at the point $t$, that is, if $V_{mn}(t) = 0$, $V_m(t) = 0$ and $V(t) = 0$, then $B(t) = 0$. 
Since $B(u) \geq 0$ for any other point $u$, we have that the left-hand side is positive semidefinite. Now let us assume that $P > 0$.  
We then have to assure that $f(t) - f(u) \geq 0$ for any point $u$, which is possible only if $t$ is maximum of the function $f$. 
The only maximum of $f(t)$ is at $t=0$ where $f(0) = 1$. Similarly if $P<0$, we must have that $t$ is a minimum, 
and the only minimum is at $t=\infty$ where $f(\infty) = -1$. 

We have now shown the following vanishing theorem. If $P>0$ then the BPS equations have no 
solutions except for $t=0$, and if $P<0$ they have no solutions except for $t=\infty$.  

Finally we notice that when BPS equation are satisfied, we have that $\F_{5i} = 0$ so that 
\bea
P &=&  \frac{1}{4} \int d^5 x \sqrt{G} \epsilon^{ijkl} \Tr\bigg[\F_{ij} \F_{kl}\bigg]v_5
\eea
If the length of the fiber is unity, we can trivially integrate over the fiber and reduce this to an integral over the contact four space. Furthermore, we notice that $2 \nabla_i v_5 = w_{i5}$ and $\nabla_5 v_5 = 0$ by the Killing equation, but in our coordinate system we in fact have that $\partial_5 v_5 = 0$ since no metric component depends on $x^5$ explicitly, this being an isometry direction, and we have that $0 = v^m w_{mi} = w_{5i}$ by picking $v^m = \delta^m_5$. 
We then notice that (after Wick rotation) $\F_{ij} = F_{ij} + i [\phi_i,\phi_j] + i (D_i \phi_j - D_j \phi_i)$ and then by using integrations by parts and the Bianchi identity and $\nabla_i v_5$, it is easy to show that 
\bea
P &=& \frac{1}{4} \int d^4 x \sqrt{g} \epsilon^{ijkl} \Tr\bigg[F_{ij} F_{kl}\bigg]
\eea
where we have also noted that $v_5 = G_{5m} v^m = G_{55} = 1$ in our choice of coordinates. Hence we again find the same BPS bound as in (\ref{BPSbound}).

\subsection{The BPS equations revisited}
We have learned that $\varphi = 0$, and that by the above vanishing theorem, there are three points we need to analyze, $\alpha = 0$ and $\alpha =\pm1$. The remaining 
case $\alpha = \infty$ is related to $\alpha=0$ by $\mb{Z}_2$ symmetry.

For the case $\alpha = 0$ the BPS equations reduce to 
\bea
D^i \phi_i &=& 0\cr
\phi_{ij} &=& 0\cr
F_{ij} + i[\phi_i,\phi_j] &=& 0\cr
F_{5i} &=& 0\cr
\phi_{5i} &=& 0
\eea
Except the first equation, the remaining are in general solved by the complex flat connection leading to
\bea
A_m &=& -i g^{-1} \partial_m g\cr
\phi_m &=& -g^{-1} h^{-1}( \partial_m h) g
\eea 
where $g=e^{i \alpha(x)}$ and $h=e^{\beta(x)}$ with $\alpha(x)$ and $\beta(x)$ being hermitian Lie algebra 
valued functions. Then  $\varphi=0$ requires that $\beta(x)$ is independent of $x^5$ coordinate. 
Then the first equation implies that 
\bea
\nabla^i (h^{-1} \nabla_i h)=0
\label{2nd}
\eea

If 
$R_{ij}$ is positive definite, one also finds $\phi_i=0$. Hence the most general BPS solutions when $\alpha = 0$ with $R_{ij}$ positive definite, are flat 
gauge connections.

Let us now turn to the case $\alpha = \pm1$. Then the BPS equations read\footnote{We note that the second and third of these equations seem to agree with eqs (3.37) and (3.38) in \cite{Unsal:2006qp}.}
\bea
D^i \phi_i &=& 0\cr
F_{ij} + i [\phi_i,\phi_j] &=& \pm \frac{1}{2} \epsilon_{ijkl} \(F^{kl} + i[\phi^k,\phi^l]\)\cr
\phi_{ij} &=& \pm \frac{1}{2} \epsilon_{ijkl} \phi^{kl}\cr
\phi_{5i} &=& 0\cr
F_{5i} &=& 0
\eea
together with the Gauss law constraint
\bea
w^{ij} \( F_{ij} + i [\phi_i,\phi_j] \)=0
\eea
As before assuming $R_{ij}$ positive definite,  one has
$\phi_i = 0$. Choose a gauge $A_5=0$. Then the last equation implies $\L_v A_i = 0$.
Then the second equation reduces to the selfdual instanton equation
\bea
F_{ij} &=& \pm \frac{1}{2} \epsilon_{ijkl} F^{kl}
\eea
which is translation invariant along the Killing direction. Of course these are further subject to the condition $w^{ij}F_{ij}=0$. 

Let us summarize: If $R_{ij}$ is positive definite, 
the most general BPS solutions are either flat gauge connections, or contact instantons subject to the condition $w^{ij}F_{ij}=0$. 
These are both required to be translationally invariant along the Killing direction in the $A_m v^m=0$ gauge.

\section{Discussion}
We would here like to speculate on possible applications of our twisted SYM theory. 

One question one might try to address is whether our theory when it is put on $S^5$, is equivalent with other SYM theory that was obtained in \cite{Hosomichi:2012ek}. One might try to answer this question by computing the partition function and then one can compare the results. 

The 5d theory is expected to have an S-duality. To illustrate this S-duality by a concrete example, which has also been considered previously in the literature (see for instance \cite{Kim:2013nva,Kim:2012tr}) we may consider $S^1 \times S^5$ where $S^5$ is a K-contact manifold and is a Hopf circle-bundle of $\mb{CP}^2$. In other words, we have a two-torus fibered over $\mb{CP}^2$. In the Euclidean case, we can consider dimensional reduction along either circle fiber and obtain two different 5d theories. Either we can obtain a 5d theory on $S^5$, or we can obtain a 5d theory on $S^1 \times \mb{CP}^2$ with an additional graviphoton term measuring the twisting of the Hopf circle-bundle along whose fiber we reduce the 6d theory on $S^1 \times S^5$ down to $S^1 \times \mb{CP}^2$. The conjectured S-duality would now say that these two theories are equivalent. Optimistically, the $SO(4)$ twisted topological field theory that we construct here can be used for better understanding this S-duality in the 5d sense.

\section*{Acknowledgement}
We would like to thank Hee-Joong Chung for helpful discussions.
DB was
supported by 2014 Research Fund of University of Seoul. 

\newpage

\appendix
\section{Spinor conventions}\label{spinor}
We choose the following representation for the gamma matrices
\bea
\Gamma^0 &=& \delta^{\alpha}_{\beta} \varepsilon^I{}_J \delta^i_j\cr
\Gamma^m &=& (\gamma^m)^{\alpha}{}_{\beta} (\sigma^1)^I{}_J \delta^i_j\cr
\Gamma^A &=& \delta^{\alpha}_{\beta} (\sigma^3)^I{}_J (\gamma^A)^i{}_j
\eea
Here $\gamma^m$ and $\gamma^A$ denote $SO(5)$ gamma matrices, and we use the convention that $\varepsilon^+{}_- = 1$, $\varepsilon^-{}_+ = -1$ and $\varepsilon^+{}_+ = 0 = \varepsilon^-{}_-$. The 6d chirality matrix is 
\bea
\Gamma = \Gamma^{012345} = \delta^{\alpha}_{\beta} (\sigma^3)^I{}_J \delta^i_j
\eea
The spinor has chirality as $\psi^{\alpha + i}$ and the supersymmetry parameter has opposite chirality $\epsilon^{\alpha - i}$. The 11d charge conjugation matrix is 
\bea
C &=& C_{\alpha\beta} \epsilon_{IJ} C_{ij}
\eea
where $\epsilon_{IJ}$ is antisymmetric with $\epsilon_{+-} = 1$. We have 
\bea
C_{\alpha\beta} &=& -C_{\beta\alpha}
\eea
We define complex conjugate as 
\bea
C^{\alpha\beta} &=& (C_{\alpha\beta})^*
\eea
These have the property 
\bea
C^{\alpha\beta} C_{\beta\gamma} &=& -\delta^{\alpha}_{\gamma}
\eea
We use these to raise and lower indices as
\bea
\psi_{\alpha} &=& C_{\alpha\beta} \psi^{\beta}\cr
\psi^{\alpha} &=& \psi_{\beta} C^{\beta\alpha}
\eea
This is iterated for multi-index objects as
\bea
\psi_{\alpha\beta...} &=& C_{\alpha\gamma} C_{\beta\delta}\cdots \psi^{\gamma\delta...}
\eea
In particular we notice that
\bea
C_{\alpha\beta} &=& C_{\alpha\gamma} C_{\beta\delta} C^{\gamma\delta}
\eea
The defining property is
\bea
(\gamma^m)^T &=& C \gamma^m C^{-1}
\eea
The indices of these matrices must sit as
\bea
((\gamma^m)^T)_{\alpha}{}^{\beta} &=& -C_{\alpha\gamma} (\gamma^m)^{\gamma}{}_{\delta} C^{\delta\beta}
\eea
Complex conjugation gives
\bea
(\gamma^m)^{\alpha}{}_{\beta} &=& -C^{\alpha\gamma} ((\gamma^{m})^{\gamma}{}_{\delta})^* C_{\delta\beta}
\eea
where we used  the relation $(\gamma^m)^{T*} = \gamma^m$.

\section{The Majorana condition after the twist}\label{Majorana}
After the twist, the Majorana condition reads
\bea
(\psi^{\alpha i})^* &=& \psi_{\alpha i}
\eea
for both 6d chiralities. We can now also expand 
\bea
\psi^{\alpha i} &=& \frac{1}{2} \(\psi C^{\alpha i} + \psi_m (\gamma^m)^{\alpha i} + 
\psi_{mn} (\gamma^{mn})^{\alpha i}\)
\eea
where $C^{\alpha i}$ is the charge conjugation matrix of $SO(5)'$, and $(\gamma^m)^{\alpha i}$ are gamma matrices of $SO(5)'$. The expansion can be inverted to extract the coefficients
\bea
\psi &=& \frac{1}{2} C_{\alpha\beta} \psi^{\alpha\beta}\cr
\psi_m &=& \frac{1}{2} (\gamma_m)_{\alpha\beta} \psi^{\alpha\beta}\cr
\psi_{mn} &=& \frac{1}{4} (\gamma_{mn})_{\alpha\beta} \psi^{\alpha\beta}
\eea
By complex conjugation 
\bea
(\psi^{\alpha i})^* &=& \frac{1}{2}\(\psi^* C_{\alpha i} + \psi_m^* (\gamma^m)_{\alpha i} + 
 \psi_{mn}^* (\gamma^{mn})_{\alpha i}\)
\eea
we can read off the reality conditions of the components as
\bea
\psi^* &=& \psi\cr
\psi_m^* &=& \psi_m\cr
\psi_{mn}^* &=& \psi_{mn}
\eea

\section{Closure of abelian 6d supersymmetry variations}\label{C}
\subsection{Closure among $Q$}
We check here the on-shell closure of the supersymmetry variations. First we find that 
\bea
[\delta_{\eta},\delta_{\epsilon}] B_{m0} &=& 2i\epsilon\eta \partial_0 B_{m0} + \partial_m \Lambda_0 - \partial_0 \Lambda_m 
\eea
where the gauge parameter is 
\bea
\Lambda_m &=& 2i\epsilon\eta \(B_{m0} - \phi_m\) 
\cr
\Lambda_0 &=& 0
\eea
Next we  get
\bea
[\delta_{\eta},\delta_{\epsilon}] B_{mn} &=& 2i\epsilon\eta \(H_{0mn}^+ - \phi_{mn}\)
\eea
We require the right-hand side is equal to 
\bea
2i\epsilon\eta \partial_0 B_{mn} + \partial_m \Lambda_n - \partial_n \Lambda_n
\eea
and we get the equation of motion as the difference
\bea
H_{0mn}^+ - \phi_{mn} - \(\partial_0 B_{mn} + \partial_m (B_{n0}-\phi_n) - \partial_n  (B_{m0}-\phi_m) \) &=& 0
\eea
which we can simplify as
\ben
H_{0mn}^+ - H_{0mn} &=& 0\label{SDEOM}
\een
where we define
\bea
H_{0mn} &=& \partial_0 B_{mn} + \partial_m B_{n0} - \partial_n B_{m0}
\eea
Then if we use the definitions (\ref{SD}), the equation of motion (\ref{SDEOM}) can be expressed as
\bea
H_{0mn}^- &=& 0
\eea

When closing supersymmetry on $\psi_{mn}$ we will again need an equation of motion. We get
\bea
[\delta_{\eta},\delta_{\epsilon}]\psi_{mn} &=& 2i\epsilon\eta \partial_0 \psi_{mn}
\eea
on the equation of motion
\ben
\partial_m \psi_n - \partial_n \psi_m - \epsilon_{mn}{}^{pqr} \partial_p\psi_{qr} - 2 i \delta C_{mn} &=& 2\partial_0 \psi_{mn}\label{Cmn}
\een
where we allow ourselves for a supersymmetry variation of $C_{mn}$ (and $\delta$ denotes that supersymmetry variation with the parameter $\epsilon$ removed).

\subsection{Closure among $Q^{\vee}$}
To simplify the computation, we shall assume the self-duality $H^+_{0mn}=H_{0mn}$ here, which will be fully justified 
 later on.

We immediately get
\bea
[\delta^{\vee}_{\eta},\delta^{\vee}_{\epsilon}] \phi_m &=& 2 i \epsilon \eta \partial_0 \phi_m
\eea
without using any equation of motion.

Closing on $B_{m0}$, we find 
\bea
[\delta^{\vee}_{\eta},\delta^{\vee}_{\epsilon}] B_{m0} &=& 2i\epsilon \eta \partial_0 B_{m0} + \partial_m 
\Lambda^\vee_0 - \partial_0 \Lambda^\vee_m
\eea
where the gauge parameter is
\bea
\Lambda^\vee_m &=& 2i \epsilon \eta (
B_{m0} + \phi_m - 2v_m \varphi )
\cr
\Lambda^\vee_0 &=& 0
\eea

Closing on $\psi_m$ we get
\bea
[\delta^{\vee}_{\eta},\delta^{\vee}_{\epsilon}] \psi_m &=& 2i \epsilon \eta \partial_0 \psi_m\cr
&& +2i \epsilon \eta v_m \(-\partial_0 \chi + v^n \nabla_n \psi - 2 \nabla_n \chi^n\)
\eea
Closure is now consistent with assuming the following equation of motion
\ben
\nabla_m \psi + 2 \nabla^n \psi_{nm} + v_m w_{np} \psi^{np} - \frac{1}{2} \epsilon_{mnpqr} w^{pq} v^r \psi^n&=& \partial_0 \psi_m
\label{f0psim}
\een

We get
\bea
[\delta^{\vee}_{\eta},\delta^{\vee}_{\epsilon}] \psi_{mn} &=& 2i\epsilon \eta \partial_0 \psi_{mn}
\eea
without using any equation of motion.

\subsection{Closure between $Q$ and $Q^{\vee}$}
Without using equations of motion, we get
\bea
[\delta_{\eta}^{\vee},\delta_{\epsilon}]\phi_m &=& 2i \epsilon \eta \L_v \phi_m
\eea
where 
\bea
\L_v f_m(x) &=& v^n \nabla_n f_m(x) + \frac{1}{2} w_{mn} f^n(x)
\eea
is the usual Lie derivative of the vector field $f_m(x)$.

We get
\bea
[\delta_{\eta}^{\vee},\delta_{\epsilon}] \psi &=& 2 i \epsilon \eta v^m \nabla_m \psi +
 i \epsilon \eta (\partial_0 \chi + 2 \nabla_m \chi^m - v^m \nabla_m \psi)
\eea
which closes upon the equation of motion (\ref{f0psim}).

We get
\bea
[\delta_{\eta}^{\vee},\delta_{\epsilon}] \psi_m &=& 2i \epsilon \eta \L_v \psi_m - 2i\epsilon \eta v_m \(-\partial_0 \psi + \nabla_n \psi^n\) 
\eea
which closes on the equation of motion 
\bea
-\partial_0 \psi + \nabla_m \psi^m &=& 0
\eea

We get
\bea
[\delta_{\eta}^{\vee},\delta_{\epsilon}] \psi_{mn} &=& i\epsilon\eta\(\epsilon_{mnpqr} v^r \(-\partial_0\psi^{pq} + \nabla^p \psi^q - \frac{1}{2} w^{pq}\chi\) + 2\(\nabla_m \chi_n - \nabla_n \chi_m\)\)
\eea
where
\bea
2\(\nabla_m \chi_n - \nabla_n \chi_m\) &=& 2 v^p \nabla_{m} \psi_{pn} - 2 v^p \nabla_n \psi_{pm} + 2 w_{[m}{}^p \psi_{|p|n]}
\eea
Using the Lie derivative expressed as
\bea
\L_v \psi_{mn} &=& v^p \nabla_p \psi_{mn} + w_{[m}{}^p \psi_{|p|n]}
\eea
we get
\bea
2\(\nabla_m \chi_n - \nabla_n \chi_m\) &=& 2\L_v \psi_{mn} - 2 v^p \nabla_p \psi_{mn} +  2 v^p \nabla_{m} \psi_{pn} - 2 v^p \nabla_n \psi_{pm} 
\eea
We then further get
\bea
&& [\delta_{\eta}^{\vee},\delta_{\epsilon}] \psi_{mn} = 2i\epsilon\eta \L_v \psi_{mn} \cr
&& ~~+ i\epsilon\eta\(\epsilon_{mnpqr} v^r \(-\partial_0\psi^{pq} + \nabla^p \psi^q - \frac{1}{2} w^{pq}\chi\) - 2 v^p \nabla_p \psi_{mn} +  2 v^p \nabla_{m} \psi_{pn} - 2 v^p \nabla_n \psi_{pm} \)
\eea
This closes on the equation of motion
\bea
-\epsilon_{mnrab} \nabla^r \psi^{mn} - 2\partial_0 \psi_{ab} + \nabla_a \psi_b - \nabla_b \psi_a - w_{ab} \chi &=& 0
\eea
from which we deduce that
\bea
C_{mn} &=& \frac{1}{2} w_{mn} \varphi
\eea
by matching with the equation of motion (\ref{Cmn}).

\end{document}